\documentclass[prb,letterpaper,aps,floatfix,superscriptaddress,twocolumn]{revtex4-1}
\usepackage{graphicx}
\usepackage{amsmath}
\usepackage{subfigure}
\newcommand{\beq}{\begin{equation}}
\newcommand{\eeq}{\end{equation}}
\newcommand{\bk}{{{\bf{k}}}}

\newcommand{\cK}{{{\cal K}}}

\newcommand{\bA}{{\bf{A}}}

\newcommand{\bq}{{\bf{q}}}

\newcommand{\bb}{{\bf{b}}}

\newcommand{\bj}{{\bf j}}
\newcommand{\bx}{{\bf x}}
\newcommand{\beqa}{\begin{eqnarray}}
\newcommand{\eeqa}{\end{eqnarray}}
\newcommand{\pdg}{{\vphantom \dag}}
\newcommand{\dg}{{\dag}}
\newcommand{\bnabla}{{\boldsymbol \nabla}} 
\newcommand{\bsigma}{{\boldsymbol \sigma}}
\newcommand{\bgamma}{{\boldsymbol \gamma}}
\newcommand{\btau}{{\boldsymbol \tau}}

\newcommand{\bGamma}{{\boldsymbol \Gamma}}
\newcommand{\upa}{\uparrow}
\newcommand{\da}{\downarrow} 
\newcommand{\ra}{\rightarrow} 
\newcommand{\cG}{{\cal G}}

\begin{document}
\title{Anomalous Hall effect in Weyl superconductors}
\author{G. Bednik}
\address{Department of Physics and Astronomy, University of Waterloo, Waterloo, Ontario 
N2L 3G1, Canada} 
\author{A.A. Zyuzin}
\address{Department of Physics, University of Basel, Klingelbergstrasse 82, CH-2056 Basel, Switzerland}
\address{A.F. Ioffe Physical-Technical Institute, 194021 Saint Petersburg, Russia}
\author{A.A. Burkov}
\address{Department of Physics and Astronomy, University of Waterloo, Waterloo, Ontario 
N2L 3G1, Canada} 
\address{ITMO University, Saint Petersburg 197101, Russia}
\date{\today}
\begin{abstract}
We present a theory of the anomalous Hall effect in a topological Weyl superconductor
with broken time reversal symmetry. 
Specifically, we consider a ferromagnetic Weyl metal with two Weyl nodes of opposite chirality near the Fermi energy. 
In the presence of inversion symmetry, such a metal experiences a weak-coupling Bardeen-Cooper-Schrieffer (BCS) instability, 
with pairing of parity-related eigenstates. Due to the nonzero topological charge, carried by the Weyl nodes, such a superconductor is necessarily topologically nontrivial, with Majorana surface states coexisting with the Fermi arcs of the normal Weyl metal. 
We demonstrate that, surprisingly, the anomalous Hall conductivity of such a superconducting Weyl metal coincides with that of 
a nonsuperconducting one, under certain conditions, in spite of the nonconservation of charge in a superconductor. 
We relate this to the existence of an extra (nearly) conserved quantity in a Weyl metal, the chiral charge. 
\end{abstract}
\maketitle
\section{Introduction}
\label{sec:1}
The focus in the explosively growing field of nontrivial electronic structure topology has recently shifted from 
gapped insulators~\cite{Hasan10,Qi11} to gapless metallic systems. 
This shift, partially anticipated in earlier work,~\cite{Volovik88,Volovik03,Haldane04,Volovik07,Murakami07}
was initiated by the theoretical discovery of Weyl semimetals,~\cite{Wan11,Ran11,Burkov11-1,Burkov11-2,Xu11}
which, along with closely related Dirac semimetals,~\cite{Kane12,Fang12,Fang13} have now been realized experimentally.~\cite{HasanTaAs,Neupane14,DingTaAs2,DingTaAs,Lu15}

Weyl points act as monopole sources of Berry curvature in momentum space. 
As a result, a Fermi surface sheet, enclosing an individual Weyl node, acquires a topological 
invariant, the Chern number, given by the flux of the Berry curvature through the Fermi surface sheet. 
This leads to both spectroscopic manifestations in the form of Fermi arc surface states,~\cite{Wan11,Haldane14}
and manifestations in electromagnetic response, in particular the anomalous Hall effect~\cite{Burkov13,Burkov14-1,Burkov14-2,Felser16}
and the longitudinal negative magnetoresistance.~\cite{Spivak12,Burkov_lmr_prl,Burkov_lmr_prb,Ong_anomaly}
Such topological response is of particular interest since it represents a manifestation of the coherent quantum mechanical behavior 
of the electrons in a given material on macroscopic scales. 

Macroscopic quantum coherent behavior also often results from electron-electron interactions and particularly interesting phenomena may arise when nontrivial electronic structure topology and electron-electron interactions work in unison, 
a prime example being the fractional quantum Hall effect. 
This motivates one to consider the issue of the interplay of electron-electron interactions and the momentum-space monopole 
topology in Weyl metals. 

In this paper we look at the electromagnetic response of a Weyl superconductor,~\cite{Meng12,Moore12,Aji14,Tanaka15,Bednik15,FanZhang14,FanZhang15,YiLi15,Ting16} namely a Weyl metal that has become 
superconducting. 
Of particular interest in this context are magnetic Weyl metals, since in this case superconductivity is automatically 
topologically nontrivial, with nodes in the gap function and Majorana surface states coexisting with the Fermi arcs of the 
Weyl metal. 
Thus the question of the anomalous Hall effect (AHE) in a topological Weyl superconductor arises naturally. 

AHE in topological superconductors with broken time reversal symmetry has been studied extensively in recent 
years.~\cite{Murakami03,Yakovenko07,Goryo08,Lutchyn08,Kallin12,Kitagawa13,Hughes10,Roy14,Lian15}
Since particle number conservation is violated in the superconducting state, one finds that the anomalous Hall conductivity is 
generally nonuniversal, depending on the nature of pairing, impurities, etc. 
In contrast, we find that in a superconducting Weyl metal with broken time reversal symmetry, the anomalous Hall conductivity 
retains the universal form of the normal Weyl metal, being proportional to the internode distance in momentum space. 
We demonstrate that this is a consequence of the emergence of a new conserved quantity in a Weyl metal, the chiral charge. 
Unlike the total charge, the conservation of chiral charge is unaffected by superconductivity, which protects the universality 
of the Hall conductivity. 

The rest of the article is organized as follows. 
In Section~\ref{sec:2} we introduce a model of a superconducting Weyl metal we will use and recap its main 
properties. 
In Section~\ref{sec:3} we evaluate the anomalous Hall conductivity of the superconducting Weyl metal by coupling 
the electrons to electromagnetic field and integrating out both the electron fields and the phase of the superconducting 
order parameter. 
This gives an induced action for the electromagnetic field only, from which the anomalous Hall conductivity may be read off. 
We conclude in Section~\ref{sec:4} with a discussion of our results and conclusions. 
\section{Weyl superconductor with broken time reversal symmetry}
\label{sec:2}
We start from the multilayer model of a magnetic Weyl semimetal, introduced in Ref.~\onlinecite{Burkov11-1}. 
The advantage of this model is its simplicity, so that all of the calculations may be done analytically. 
At the same time, this model captures all of the essential properties of a generic Weyl semimetal with broken time reversal 
symmetry. 

The momentum space Hamiltonian of the magnetic multilayer structure has the form
\beq
\label{eq:1}
H_0 = v_F \tau^z (\hat z \times \bsigma) \cdot \bk + \hat t(k_z) + b \sigma^z. 
\eeq
Here $\hat z$ is the growth direction of the multilayer, the eigenstates of $\tau^z$ describe the top and bottom surfaces of the 
topological insulator (TI) layers, $\bsigma$ are Pauli matrices, describing the real spin degree of freedom and the $b \sigma^z$ 
term corresponds to the exchange spin splitting in the ferromagnetic states of the multilayer. 
$\hbar = 1$ units are used in Eq.~\eqref{eq:1} and in the rest of the paper. 
The operator $\hat t(k_z)$ is given by
\beq
\label{eq:2}
\hat t(k_z) = t_S \tau^x + \frac{t_D}{2}( \tau^+ e^{i k_z d} + h.c. ), 
\eeq
and describes the motion of the electrons in the growth direction of the multilayer. 
$t_{S,D} > 0$ are amplitudes, corresponding to tunneling between surface states within the same (S) or 
neighboring (D) TI layers, and $d$ is the superlattice period in the growth direction. 

The eigenvalues of Eq.~\eqref{eq:1} correspond to four bands, two of which touch at a pair of Weyl points, separated 
along the $z$-axis in momentum space. We will assume that the Fermi energy $\epsilon_F$ is close to the location 
of the Weyl nodes, which is exactly at zero energy due to particle-hole symmetry of the model. 
While this particle-hole symmetry is generally not present, it does characterize low-energy states near the Dirac point, 
which occurs in our model when $t_S = t_D$ at $\bk = (0,0,\pi/d)$. 
In what follows we will assume that $|t_S - t_D|$ is small, while $b \ll t_S$, i.e. the important low-energy states are not far 
from the Dirac point. 
In this case, particle-hole symmetry may be assumed to be approximately present. 

For the subsequent discussion of superconductivity in this model, it is convenient to partially diagonalize Eq.~\eqref{eq:1}
in order to separate the pair of bands that touch at the Weyl nodes from the other pair. 
This separation is accomplished by the canonical transformation
\beq
\label{eq:3}
\sigma^{\pm} \rightarrow \tau^z \sigma^{\pm}, \,\, \tau^{\pm} \rightarrow \sigma^z \tau^{\pm}, 
\eeq
which gives
\beq
\label{eq:4}
H_0 = v_F (\hat z \times \bsigma) \cdot \bk + \hat m(k_z) \sigma^z, 
\eeq
where 
\beq
\label{eq:5}
\hat m(k_z) = b + \hat t(k_z). 
\eeq
The operator $\hat m(k_z)$ may now be diagonalized separately, which accomplishes the desired separation into 
high- and low-energy states. 
Namely, the eigenvalues of $\hat m(k_z)$ have the form
\beq
\label{eq:6}
m_r(k_z) = b + r t(k_z) \equiv b + r \sqrt{t_S^2 + t_D^2 + 2 t_S t_D \cos(k_z d)}, 
\eeq 
where $r = \pm$. The two low-energy bands correspond to $r = -$. 
They touch at two Weyl points, whose location is given by the solution of the equation $m_-(k_z)  = 0$. 
The $4 \times 4$ Hamiltonian $H_0$ separates into two independent $2 \times 2$ blocks $H_{0 r}$. 

We will assume that the pairing interaction has the simplest local $s$-wave form and is diagonal 
in the band index $r$, corresponding to pairing of parity-related eigenstates (this assumption is physically reasonable and does 
make calculations analytically tractable, but is not essential as far as the final results are concerned)
\beq
\label{eq:7}
H_{int} = - U \int d^3 x \sum_{r r'} c^\dg_{r \upa} (\bx) c^\dg_{r \da} (\bx) c^\pdg_{r' \da} (\bx) c^\pdg_{r' \upa}(\bx).
\eeq
Decoupling the pairing interaction in the standard BCS mean-field approximation, we obtain
\beqa
\label{eq:8}
H&=&\sum_{r \bk} \left[v_F (\hat z \times \bsigma) \cdot \bk + m_r(k_z) \sigma^z - \epsilon_F \right] c^\dg_r(\bk) c^\pdg_r(\bk) \nonumber \\
&+&\Delta c^\dg_{r \upa}(\bk) c^\dg_{r \da}(-\bk) + \Delta^\dg c^\pdg_{r \da}(-\bk) c^\pdg_{r \upa}(\bk),
\eeqa
where 
\beq
\label{eq:9}
c^\dg(\bx) = \frac{1}{\sqrt{V}} \sum_\bk c^\dg(\bk) e^{- i \bk \cdot \bx}. 
\eeq
As evident from Eq.~\eqref{eq:8} we will assume that the pairing occurs in the BCS channel, i.e. 
the states with opposite momenta $\bk$ and $-\bk$ are paired. These states are related by the exact inversion
symmetry, that we assume to be present in our system. For this, and also for phase space reasons, the BCS 
pairing state is much more likely to be realized in our system than other kinds of superconducting states, as discussed in detail
in Ref.~\onlinecite{Bednik15}. 

\begin{figure}[t]
  \includegraphics[width=8cm]{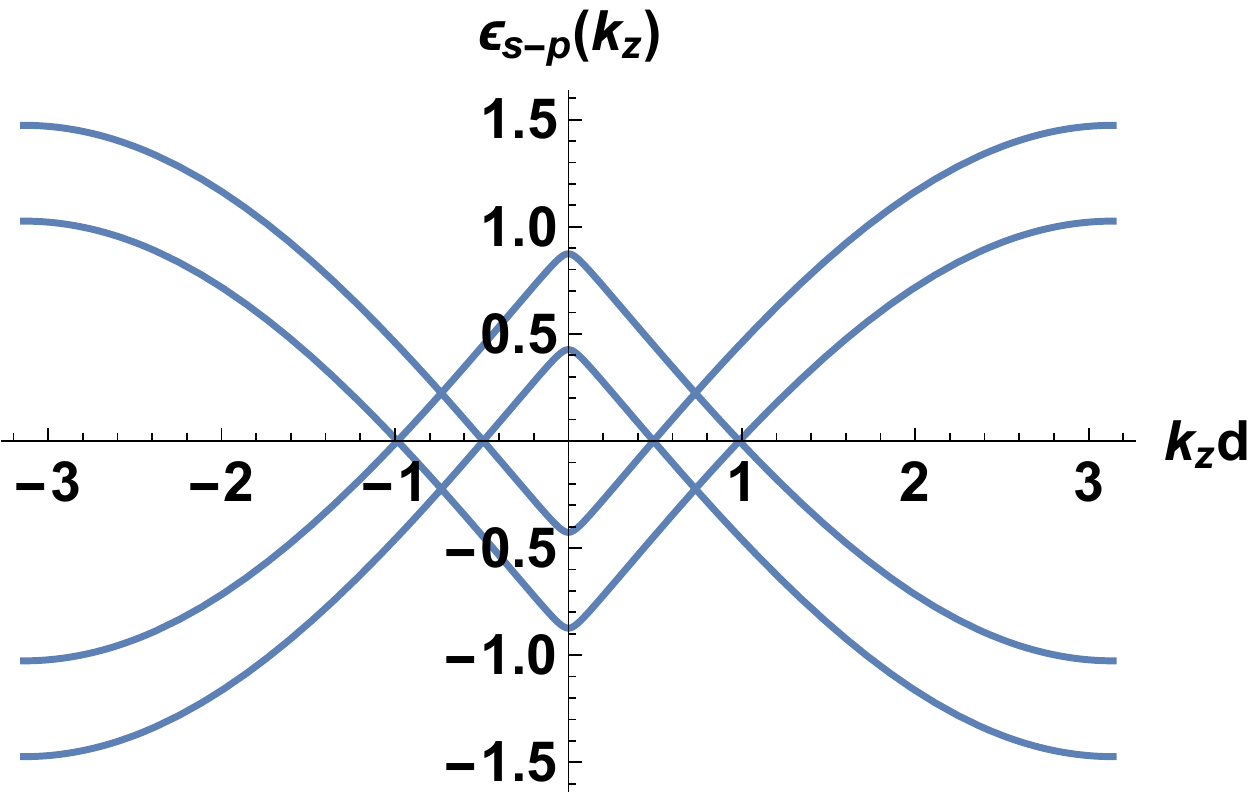}
  \caption{Plot of the band dispersion $\epsilon_{s - p}(\bk)$ along the $z$-axis in momentum space for 
  $t_S = 1, t_D = 0.95, b = 0.7, \epsilon_F= 0.2, \Delta = 0.1$. The zero of $k_z$ has been shifted to $k_z = \pi/d$ for
  presentation convenience.} 
  \label{fig:1}
\end{figure}

To find the eigenstates of the BCS Hamiltonian we introduce the following Nambu spinor notation
\beqa
\label{eq:10}
\Psi(\bk)&=&[c^\pdg_\upa(\bk), c^\pdg_\da(\bk), c^\dg_\da(-\bk), c^\dg_\upa(-\bk)] \nonumber \\
&\equiv& [\Psi_{1 \upa}(\bk), \Psi_{1 \da}(\bk), \Psi_{2 \upa}(\bk), \Psi_{2 \da}(\bk)],
\eeqa
where the indices $1,2$ label the particle and hole components of the Nambu spinor correspondingly. 
Taking $\Delta$ to be real, and introducing Pauli matrices $\btau$ that act on the particle-hole indices, 
we obtain
\beq
\label{eq:11}
H_r = v_F (\hat z \times \bsigma) \cdot \bk + m_r(k_z) \sigma^z - \epsilon_F \tau^z + \Delta \tau^x \sigma^z. 
\eeq
The eigenvalues of Eq.~\eqref{eq:11} are given by
\beqa
\label{eq:12}
&&\epsilon_{s r p}(\bk) \equiv s \epsilon_{r p}(\bk) \nonumber \\
&= &s \sqrt{\epsilon^2_r(\bk) + \epsilon_F^2 + \Delta^2 + 2 p 
\sqrt{\epsilon^2_r(\bk) \epsilon_F^2 + \Delta^2 m^2_r(k_z)}}, \nonumber \\
\eeqa
where $s, p = \pm$ and $\epsilon_r(\bk) = \sqrt{v_F^2 (k_x^2 + k_y^2) + m^2_r(k_z)}$. 
The two particle-hole antisymmetric states, corresponding to $r = -$ and $p = -$ touch at 
four points along the $z$-axis in momentum space, given by the equation
\beq
\label{eq:13}
m_-(k_z) = \pm \sqrt{\epsilon_F^2 + \Delta^2}. 
\eeq
The superconducting gap thus has four nodes, which is inevitable due to the nonzero flux of the Berry curvature 
through the two Fermi surface sheets, each enclosing a Weyl point,~\cite{YiLi15} see Fig.~\ref{fig:1}. 
In close analogy to the normal Weyl semimetal, the four nodes may be regarded as transition points in momentum space, 
at which topology of the Hamiltonian Eq.~\eqref{eq:11} changes. 
Correspondingly, there exist Majorana arc surface states, which connect projections of the node locations onto the 
surface Brillouin zone (BZ),~\cite{Meng12,Tanaka15,Bednik15,YiLi15} i.e. Eq.~\eqref{eq:11} describes a topological superconductor 
with broken time reversal symmetry. 

\section{Anomalous Hall conductivity of the Weyl superconductor}
\label{sec:3}
To describe the electromagnetic response of the Weyl superconductor we couple electrons to electromagnetic field 
and integrate out the fermionic degrees of freedom to obtain an induced action for the electromagnetic field only. 
In a superconductor this procedure is complicated by the fact that not only the bare electronic degrees of freedom need 
to be integrated out, but also the emergent degrees of freedom of the superconducting state, in particular the phase of the superconducting order parameter (amplitude fluctuations are massive and may be ignored). 
We start from an imaginary time action 
\beqa
\label{eq:14}
S&=&\int_0^{\beta} d \tau \int d^3 x \sum_r \left[c^\dg_r(\bx, \tau) (\partial_{\tau} + H_{0r} - \epsilon_F) c^\pdg_r(\bx, \tau)\right. \nonumber \\
&+&\left. \Delta(\bx, \tau) c^\dg_{r \upa}(\bx, \tau) c^\dg_{r \da}(\bx, \tau) + \Delta^\dg(\bx, \tau) c^\pdg_{r \da}(\bx, \tau) 
c^\pdg_{r \upa}(\bx, \tau) \right. \nonumber \\
&+& \left.\frac{1}{U} |\Delta(\bx, \tau)|^2\right], 
\eeqa
where the crystal momentum $\bk$ in Eq.~\eqref{eq:1} has been replaced by $-i \bnabla$ (we will ignore the distinction between continuum and lattice gradients here as it will not be important in what follows). 
We couple the electromagnetic field to the electrons by making a replacement
\beq
\label{eq:15}
- i \bnabla \ra -i \bnabla + e \bA(\bx, \tau) ,\,\, \partial_{\tau} \ra \partial_\tau + e A_0(\bx, \tau), 
\eeq
and perform a gauge transformation 
\beq
\label{eq:16}
c(\bx, \tau) \ra c(\bx, \tau) e^{i \theta(\bx, \tau)}, 
\eeq
where $2 \theta(\bx, \tau)$ is the phase of the superconducting order parameter $\Delta(\bx, \tau)$. 
We will take $|\Delta(\bx, \tau)| \equiv \Delta$ to be constant, as fluctuations of the magnitude of the order parameter do not affect 
the electromagnetic response qualitatively. 
For notational convenience we will introduce new variables
\beq
\label{eq:17}
\tilde A_{\mu} = A_{\mu} + \frac{1}{e} \partial_{\mu} \theta, 
\eeq
since the electromagnetic fields and the derivatives of the phase of the superconducting order parameter always 
enter the action in such combinations due to gauge invariance. 

Passing to the Nambu notation and Fourier transforming, we obtain
\beqa
\label{eq:18}
S&=&\sum_{\bk, i \omega} \sum_{\bk', i \omega'} \sum_r \left[\cG_{0 r}^{-1}(\bk, i \omega) \delta_{\bk \bk'} 
\delta_{\omega \omega'} + \delta \cG_r^{-1}(\bk, i \omega | \bk', i \omega') \right] \nonumber \\
&\times&\Psi^\dg_r(\bk, i \omega) \Psi^\pdg_r(\bk', i\omega'), 
\eeqa
where 
\beq
\label{eq:19}
\cG^{-1}_{0 r}(\bk, i \omega) = -i \omega + v_F(\hat z \times \bsigma) \cdot \bk + m_r(k_z) \sigma^z - \epsilon_F \tau^z + 
\Delta \tau^x \sigma^z, 
\eeq
and 
\beqa
\label{eq:20}
&&\delta \cG^{-1}_r(\bk, i \omega | \bk', i \omega') = \frac{e}{\sqrt{V \beta}}\left[\tilde A_0(\bk - \bk', i \omega - i \omega') \right. \nonumber \\
&+&\left. \frac{1}{e} \frac{\partial H_r(\bk')}{\partial \bk'} \cdot \tilde \bA(\bk - \bk', i \omega - i \omega') \right] \tau^z. 
\eeqa
It is implicit that the sums over $\bk$ are restricted to half the momentum space to compensate for the doubling of degrees of freedom when introducing the Nambu spinor notation. 

Integrating out the Nambu fields, we obtain, at second order in the gauge fields $\tilde A_{\mu}$
\beqa
\label{eq:21}
S&=&\frac{1}{2} \sum_{\bk, i \omega} \sum_{\bk', i\omega'} \sum_r \textrm{Tr}\, \cG_{0 r}(\bk , i \omega) \delta 
\cG^{-1}_r(\bk, i \omega | \bk', i \omega') \nonumber \\
&\times& \cG_{0 r}(\bk', i \omega') \delta \cG^{-1}_r(\bk', i \omega' | \bk, i \omega), 
\eeqa
which takes the following general form
\beq
\label{eq:22}
S = \frac{1}{2} \sum_{\bq, i \Omega} \tilde \Pi_{\mu \nu} (\bq, i \Omega) \tilde A_{\mu}(\bq, i \Omega) \tilde A_{\nu}(-\bq, -i \Omega),  
\eeq  
where $\mu, \nu = 0, x, y, z$ and summation over repeated indices is implicit. 
To obtain the measurable electromagnetic response, we need to further integrate out the phase of the superconducting 
order parameter, which enters in Eq.~\eqref{eq:22}. 
Explicitly, we have
\beq
\label{eq:23}
S = S_{AA} + S_{\theta \theta} + S_{A \theta}, 
\eeq
where 
\beq
\label{eq:24}
S_{AA} = \frac{1}{2} \sum_{\bq, i \Omega} \tilde \Pi_{\mu \nu} (\bq, i \Omega) A_{\mu}(\bq, i \Omega) A_{\nu}(-\bq, -i \Omega), 
\eeq
\beq
\label{eq:25}
S_{\theta \theta} =  \frac{1}{2} \sum_{\bq, i \Omega} q_{\mu} q_{\nu} \tilde \Pi_{\mu \nu}(\bq, i \Omega) \theta(\bq, i \Omega) 
\theta(-\bq, -i \Omega), 
\eeq
with $q_0 \equiv -i \Omega$, and, finally
\beqa
\label{eq:26}
S_{A \theta}&=&\frac{i}{2} \sum_{\bq, i \Omega} \left[q_{\mu} \tilde \Pi_{\mu \nu}(\bq, i \Omega) \theta(\bq, i\Omega) 
A_{\nu}(-\bq, -i \Omega) \right.\nonumber \\
&-&\left. q_{\nu} \tilde \Pi_{\mu \nu}(\bq, i \Omega) A_{\mu}(\bq, i \Omega) \theta(-\bq, -i \Omega) \right]. 
\eeqa
Integrating out $\theta(\bq, i \Omega)$, we obtain~\cite{Lutchyn08}
\beq
\label{eq:27}
S = \frac{1}{2} \sum_{\bq, i\Omega} \Pi_{\mu \nu}(\bq, i \Omega) A_{\mu}(\bq, i \Omega) A_{\nu}(-\bq, -i \Omega), 
\eeq
where 
\beq
\label{eq:28}
\Pi_{\mu \nu} = \tilde \Pi_{\mu \nu} - \frac{q_{\mu'} q_{\nu'} \tilde \Pi_{\mu \mu'} \tilde \Pi_{\nu' \nu}}{q_{\alpha} q_{\beta}
\tilde \Pi_{\alpha \beta}}. 
\eeq
We will be interested only in the ``topological" part of the electromagnetic response, which has the appearance of a Chern-Simons 
term and thus contains components of the form $\epsilon_{i j} A_0 \partial_i A_j$ and $\epsilon_{i j} \partial_{\tau} A_i A_j$, 
where $\epsilon^{i j}$ is the fully antisymmetric tensor. 
This implies that the corresponding components of the response function $\tilde \Pi_{\mu \nu}(\bq, i \Omega)$ are of the form
\beq
\label{eq:29}
\tilde \Pi_{0 i}(\bq, i \Omega) \sim \epsilon_{ij} q_j, \,\, \tilde \Pi_{i j}(\bq, i \Omega) \sim  i \Omega \epsilon_{ij}. 
\eeq
We will restrict ourselves exclusively to the DC limit of the topological response, which is obtained by taking 
the limit of $\bq \ra 0$ first, followed by $\Omega \ra 0$.~\cite{Burkov14-1}
Taking these limits in Eq.~\eqref{eq:28} while keeping in mind Eq.~\eqref{eq:29}, we obtain
\beq
\label{eq:30}
\Pi_{0 i}(\bq, i \Omega) \approx - \frac{q_j}{i \Omega} \tilde \Pi_{i j}(\bq, i \Omega), 
\eeq
and 
\beq
\label{eq:31}
\Pi_{ij}(\bq, i \Omega) \approx \tilde \Pi_{i j}(\bq, i\Omega), 
\eeq
which means that the DC response is fully determined by the corresponding limit of the function $\tilde \Pi_{i j}(\bq, i \Omega)$. 
Note that, given Eq.~\eqref{eq:31}, Eq.~\eqref{eq:30} is dictated by gauge invariance, which requires $q_{\mu} \Pi_{\mu \nu} = 0$. 
The fact that $\Pi_{0 i}(\bq, i \Omega)$ vanishes in the DC limit reflects perfect screening of the density response to magnetic field 
by the supercurrent. 

The DC Hall conductivity, which we are interested in, may then be obtained as
\beq
\label{eq:32}
\sigma_{xy} = \lim_{i \Omega \ra 0} \frac{\tilde \Pi_{xy}(0, i \Omega)}{i \Omega}, 
\eeq
which coincides with the standard Kubo formula expression. 
Using Eqs.~\eqref{eq:19},\eqref{eq:20}, and \eqref{eq:21} we obtain
\beqa
\label{eq:33}
&&\sigma_{xy} = \frac{e^2 v_F^2}{V} \sum_{r \bk} \frac{n_F[\epsilon_{s' r p'}(\bk)] - n_F[\epsilon_{s r p}(\bk)]}
{[\epsilon_{s' r p'}(\bk) - \epsilon_{s r p}(\bk)]^2} \nonumber \\
&\times&\textrm{Im} \left[\langle z^{s r p}(\bk) | \tau^z \sigma^y | z^{s' r p'}(\bk)\rangle \langle z^{s' r p'}(\bk) | \tau^z \sigma^x | z^{s r p}(\bk)\rangle\right], 
\nonumber \\
\eeqa
where $|z^{s r p}(\bk)\rangle$ are the eigenvectors of Eq.~\eqref{eq:11} corresponding to the eigenvalues $\epsilon_{s r p}(\bk)$. 
Unlike the analogous expression for the Hall conductivity of a normal metal or an insulator, Eq.~\eqref{eq:33} can not generally 
be rewritten in terms of an integral of the Berry curvature over the occupied states due to the presence of the operator $\tau^z$, 
which acts on the Nambu particle-hole indices and reflects the nonconservation of charge in a superconductor. 

To proceed, we will first evaluate Eq.~\eqref{eq:33} in the limit $\epsilon_F = 0$, i.e. the Fermi energy in the normal state
coinciding with the Weyl nodes, since this limit is particularly simple to analyze and understand. 
We will then generalize our results to the case $\epsilon_F \neq 0$. 
\subsection{Hall conductivity at $\epsilon_F = 0$}
\label{sec:3.1}
Simplification in the $\epsilon_F = 0$ limit stems from the fact that in this case the Hamiltonian Eq.~\eqref{eq:11} breaks up into 
two independent blocks, acting onto the spin and the particle-hole pseudospin subspaces. 
Diagonalizing the pseudospin block of the Hamiltonian one obtains
\beq
\label{eq:34}
H_{r p} = v_F (\hat z \times \bsigma) \cdot \bk + [m_r(k_z) + p \Delta] \sigma^z, 
\eeq
where, as before, $p = \pm$ labels the particle-hole symmetric and particle-hole antisymmetric eigenstates. 
Viewing $H_{r p}$ as Hamiltonians describing fictitious two-dimensional systems, parametrized by $k_z$, 
there are topological phase transitions, characterized by the change of sign of the Dirac masses at 
$m_r(k_z) = \pm \Delta$.~\cite{Meng12}
The eigenvalues, and the corresponding eigenvectors, are given by
\beq
\label{eq:35}
\epsilon_{s r p}(\bk) \equiv s \epsilon_{r p}(\bk) = s \sqrt{v_F^2 (k_x^2 + k_y^2) + m_{r p}^2(k_z)}, 
\eeq
where $m_{r p}(k_z) = m_r (k_z) + p \Delta$, and 
\beqa
\label{eq:36}
|z^{s r p}(\bk)\rangle&=&\frac{1}{\sqrt{2}} \left(\sqrt{1 + s \frac{m_{r p}(k_z)}{\epsilon_p(\bk)}}, -i s e^{i \phi}\sqrt{1 - s 
\frac{m_{r p}(k_z)}{\epsilon_p(\bk)}}\right) \nonumber \\
&\otimes&\frac{1}{\sqrt{2}} (1, p), 
\eeqa
where $e^{i \phi} = k_+ /\sqrt{k_x^2 + k_y^2}$. 
Substituting these into the expression for the Hall conductivity, Eq.~\eqref{eq:33}, we obtain
\beq
\label{eq:37}
\sigma_{xy} = \frac{e^2 v_F^2}{V} \sum_{r \bk} \frac{m_{r +}(k_z) \epsilon_-(\bk) + m_{r -}(k_z) \epsilon_+(\bk)}
{\epsilon_{r +}(\bk) \epsilon_{r -}(\bk) [\epsilon_{r +}(\bk) + \epsilon_{r -}(\bk)]^2}, 
\eeq
which, upon evaluating the integral over the transverse momentum components $k_{x,y}$, gives 
\beqa
\label{eq:38}
&&\sigma_{xy} = \frac{e^2}{8 \pi^2} \sum_r \int_{-\pi/d}^{\pi/d} d k_z \left\{\left(1 - \frac{\Delta^2}{3 m_r^2(k_z)}\right) \textrm{sign}[m_r(k_z)] 
\right. \nonumber \\
&\times&\left.\Theta(|m_r(k_z)| - \Delta) + \frac{2 m_r(k_z)}{3 \Delta} \Theta(\Delta - |m_r(k_z)|) \right\}. 
\eeqa
This expression contains two distinct terms, corresponding to two physically different contributions to the Hall conductivity. 
The first term arises from the interval of $k_z$, where $|m_r(k_z)| > \Delta$.
In this interval the Fermi arcs of the nonsuperconducting Weyl semimetal still exist. 
The corresponding contribution to the Hall conductivity, however, is not simply proportional to the length of the Fermi arcs, as 
in the normal state, but contains a nonuniversal contribution, proportional to $\Delta^2$, which is to be expected as the 
superconducting state violates charge conservation.
The second term comes from the interval in which $\Delta > |m_r(k_z)|$, which we assume may only be satisfied for the $r = -$ 
pair of bands near the locations of the Weyl nodes, where the bands touch and thus $m_-(k_z) = 0$. 
In this interval Fermi arcs are replaced by Majorana arcs, which by themselves do not contribute to the Hall conductivity as 
they are neutral at zero energy. The second term thus arises from the bulk states.  

Thus, at first sight, the anomalous Hall conductivity of a Weyl superconductor is no longer universal, which is, in principle, to be expected due to violation of the charge conservation in a superconductor. 
Closer inspection of Eq.~\eqref{eq:38}, however, shows that this conclusion in the case of a Weyl superconductor is premature. 
Indeed, assuming $\Delta \ll b$, which is a physically reasonable assumption (although it may not be the case always), 
the second term clearly vanishes upon integration over $k_z$, as $m_-(k_z)$ changes sign at each Weyl point and varies linearly 
near it. 
The contribution of the term, proportional to $\Delta^2$ may be easily estimated perturbatively in $b$. 
Since at $b = 0$  the contribution of this term, along with the rest of the Hall conductivity, vanishes, the 
leading contribution is proportional to $b$ and is given by
\beq
\label{eq:39}
\sum_r \int_{-\pi/d}^{\pi/d} d k_z \frac{\Delta^2}{3 m_r^2(k_z)} \textrm{sign}[m_r(k_z)] \Theta(|m_r(k_z)| - \Delta) \sim 
\frac{\Delta^2 b}{t_S^3 d}, 
\eeq
where we have assumed, as before, that $|t_S - t_D| \ll t_S$. 
This means that this contribution vanishes if the band dispersion away from the Weyl points is taken to be strictly linear, which 
corresponds to sending $t_S$ to infinity relative to all other energy scales. 
Thus, as long as the band dispersion away from the Weyl nodes is taken to be strictly linear, the anomalous Hall 
conductivity is given by
\beqa
\label{eq:40}
\sigma_{xy}&=&\frac{e^2}{8 \pi^2} \sum_r \int_{-\pi/d}^{\pi/d} d k_z \textrm{sign}[m_r(k_z)] \Theta(|m_r(k_z)| - \Delta)
\nonumber \\
&=& \frac{e^2}{8 \pi^2} \sum_r \int_{-\pi/d}^{\pi/d} d k_z \textrm{sign}[m_r(k_z)] = \frac{e^2 \cK}{4 \pi^2}, 
\eeqa
where 
\beq
\label{eq:41}
\cK = \frac{2}{d} \arccos\left(\frac{t_S^2 + t_D^2 - b^2}{2 t_S t_D} \right), 
\eeq
is the separation between the Weyl nodes in momentum space in the normal state. 
Linearity if the band dispersion near the Weyl nodes was also assumed in going from the first to the 
second line of Eq.~\eqref{eq:40}. 

Thus, the conclusion we may draw is that, in spite of the fact that the charge conservation is violated by superconductivity, 
the Hall conductivity of a Weyl superconductor retains the universal value, characteristic of the nonsuperconducting Weyl 
semimetal, as long the the dispersion away from the Weyl nodes may be taken to be linear. 
This result is in contrast to what one would normally expect for a superconductor with broken time reversal symmetry and 
may be regarded as a manifestation of the nonrenormalization of chiral anomaly by superconductivity.~\cite{Ting16}
\subsection{Hall conductivity at $\epsilon_F > 0$}
\label{sec:3.2}
The above results may be generalized to the more natural case of a Weyl metal with a finite Fermi energy, which 
we take for concreteness to be positive. 
The calculation in this case is significantly more tedious, but the end results are still relatively simple. 
We obtain
\beq
\label{eq:42}
\sigma_{xy} = \frac{e^2 v_F^2}{V} \sum_{r \bk} m_r(k_z) \frac{1 - 4 (\epsilon_F^2 + \Delta^2)/
[\epsilon_{r +}(\bk) + \epsilon_{r -}(\bk)]^2}{\epsilon_{r +}(\bk) \epsilon_{r -}(\bk) [\epsilon_{r +}(\bk) + \epsilon_{r -}(\bk)]}. 
\eeq
After integration over $k_{x,y}$ this gives
\beqa
\label{eq:43}
\sigma_{xy}&=&\frac{e^2}{8 \pi^2}\sum_r \int_{-\pi/d}^{\pi/d} d k_z \left\{\left[1 + \frac{\Delta^2}{\epsilon_F^2} - 
\frac{\Delta^2 |m_r(k_z)|}{2 \epsilon_F^3} \right. \right. \nonumber \\
 &\times&\left. \left. \log\left(\frac{|m_r(k_z)| + \epsilon_F}{|m_r(k_z)| - \epsilon_F} \right) \right]
\textrm{sign}[m_r(k_z)] \right. \nonumber \\
 &\times& \left.\Theta\left[|m_r(k_z)| - \sqrt{\epsilon_F^2 + \Delta^2}\right] \right. \nonumber \\
&+&\left.\frac{m_r(k_z)}{\epsilon_F} \left[\sqrt{1 + \frac{\Delta^2}{\epsilon_F^2}} - \frac{\Delta^2}{2 \epsilon_F^2} 
\log\left(\frac{\sqrt{\epsilon_F^2 + \Delta^2} + \epsilon_F}{\sqrt{\epsilon_F^2 + \Delta^2} - \epsilon_F}\right)\right] \right. \nonumber \\
&\times&\left.\Theta\left[\sqrt{\epsilon_F^2 + \Delta^2} - |m_r(k_z)|\right] \right\}, 
\eeqa
which reduces to Eq.~\eqref{eq:38} in the limit $\epsilon_F \ra 0$. 
Assuming $\epsilon_F \ll t_S$, one finds, just as in the $\epsilon_F =0$ case, that the leading correction to the normal-state 
Hall conductivity is $O(e^2 \Delta^2 b / t_S^3 d)$ and is thus negligible. 
Thus we obtain, taking $\Delta \ra 0$ in Eq.~\eqref{eq:43}
\beqa
\label{eq:44}
\sigma_{xy}&=&\frac{e^2}{8 \pi^2} \sum_r \int_{-\pi/d}^{\pi/d} d k_z \textrm{sign}[m_r(k_z)] \Theta[|m_r(k_z)| - \epsilon_F] 
\nonumber \\
&+&\frac{m_r(k_z)}{\epsilon_F} \Theta[\epsilon_F - |m_r(k_z)|], 
\eeqa
which coincides with the known result for the anomalous Hall conductivity of a normal Weyl metal.~\cite{Burkov14-1}
Moreover, as first explained in Ref.~\onlinecite{Burkov14-1}, as long as $\epsilon_F \ll b \ll t_S$, correction to $\sigma_{xy}$ 
from a nonzero Fermi energy is also negligibly small, being $O(e^2 \epsilon_F b /t_S d)$. 
Thus, we arrive as the conclusion that the anomalous Hall conductivity of a superconducting Weyl metal retains a universal 
form of Eq.~\eqref{eq:40}, with small corrections, related to nonlinearity of the band dispersion near the Weyl nodes. 
\section{Discussion and conclusions}
\label{sec:4}
The result we have obtained, that the anomalous Hall conductivity of a superconducting Weyl metal 
is approximately unaffected by superconductivity, appears, at first sight, rather strange. 
Indeed, universality of the Hall conductivity in a gapped quantum Hall insulator, or in a Weyl semimetal, 
depends crucially on charge conservation. 
This may be seen, for example, from Eq.~\eqref{eq:38}.
The integrand in Eq.~\eqref{eq:38} at a fixed value for $k_z$ gives the Hall conductivity of a two-dimensional (2D)
quantum Hall insulator, in the presence of superconductivity. 
There is clearly a nonzero correction, even when $\Delta < |m|$, i.e. even before the transition to a 2D topological 
superconductor. This expresses the fact that even in gapped quantum Hall insulator, the Hall conductivity 
becomes nonquantized in the presence of superconductivity, due to the violation of charge conservation. 
However, in the 3D Weyl semimetal case, these corrections cancel each other out when the integration over $k_z$ is 
performed and the Hall conductivity remains the same as in the normal state. 
This must be a consequence of some remaining symmetry, or a conservation law, which is not violated by superconductivity and which protects the universal value of the Hall conductivity of a Weyl semimetal. 
This conservation law is the {\em conservation of chiral charge}. 

Indeed, the electromagnetic response of a nonsuperconducting Weyl semimetal may be described by a 
topological term in the induced action for the electromagnetic field~\cite{Zyuzin12-1,Hughes15}
\beqa
\label{eq:45}
S&=&\frac{e^2}{32 \pi^2} \int dt d^3 x \theta(\bx) \epsilon^{\mu \nu \alpha \beta} F_{\mu \nu} F_{\alpha \beta} \nonumber \\
&=&- \frac{e^2}{8 \pi^2} \int dt d^3 x \partial_{\mu} \theta(\bx) \epsilon^{\mu \nu \alpha \beta} A_{\nu} \partial_{\alpha} A_{\beta}, 
\eeqa
where $\theta(\bx) = 2 \bb \cdot \bx$, and $\bb$ is the vector, describing the momentum-space separation between the 
Weyl nodes (restricting ourselves to only a single pair of nodes for simplicity). 
An immediate consequence of Eq.~\eqref{eq:45} is the anomalous Hall effect, with a universal conductivity, determined 
only by the vector $\bb$
\beq
\label{eq:46}
j_{\nu} = \frac{e^2}{2 \pi^2} b_{\mu} \epsilon^{\mu \nu \alpha \beta} \partial_{\alpha} A_{\beta}. 
\eeq
On the other hand, Eq.~\eqref{eq:45} is equivalent to the anomalous conservation law for the chiral charge, or the 
chiral anomaly
\beq
\label{eq:47}
\partial_{\mu} j_5^{\mu} = \frac{e^2}{16 \pi^2} \epsilon^{\mu \nu \alpha \beta}F_{\mu \nu} F_{\alpha \beta}, 
\eeq
where 
\beq
\label{eq:48}
j_5^0 = \psi^\dg_R \psi^\pdg_R - \psi^\dg_L \psi^\pdg_L, 
\eeq
is the chiral charge,
\beq
\label{eq:49}
\bj_5 = \psi^\dg_R \gamma^0 \bgamma \psi^\pdg_R - \psi^\dg_L \gamma^0 \bgamma \psi^\pdg_L, 
\eeq
 is the chiral current, the $R,L$ indices refer to the right- and left-handed Weyl fermions, and $\gamma^{\mu}$ are 
the corresponding Dirac $\gamma$-matrices. 
Eq.~\eqref{eq:47} holds as long as the chiral charge is conserved (in the absence of the electromagnetic field), 
i.e. as long as the action is invariant with respect to the transformation
\beq
\label{eq:50}
\psi_R \ra \psi_R e^{i \theta}, \,\, \psi_L \ra \psi_L e^{- i \theta}. 
\eeq
Importantly, BCS superconductivity, with pairing of parity-related eigenstates, does not violate the chiral symmetry, 
expressed by Eq.~\eqref{eq:50}, even though it does violate the ordinary charge conservation symmetry
\beq
\label{eq:51}
\psi_R \ra \psi_R e^{i \theta}, \,\, \psi_L \ra \psi_L e^{i \theta}. 
\eeq
This connection between the Hall conductivity and the chiral anomaly in a Weyl semimetal guarantees 
that the Hall conductivity is unaffected by superconductivity, as long as the chiral charge may be regarded as a 
conserved quantity. When is the chiral charge conserved? 

In the condensed matter context, chiral symmetry, or chiral charge conservation, is an emergent property of Dirac band-touching points at time-reversal invariant momenta (TRIM) in the first BZ. 
Indeed, the most general time-reversal and parity-invariant momentum-space Hamiltonian, describing a system with four low-energy states, i.e. two spin and two orbital states (this is the minimal number of states, necessary 
to realize a Dirac point) may be written as~\cite{Fu-Kane}
\beq
\label{eq:52}
H(\bk) =d_0(\bk) I + \sum_{a = 1}^5 d_a(\bk) \Gamma^a, 
\eeq
where $\Gamma^a$ are the five matrices, realizing the Clifford algebra $\{\Gamma^a, \Gamma^b \} = 2 \delta^{ab}$, even under the product of parity and time reversal $P \Theta$. 
The five $\Gamma$-matrices are given by~\cite{Fu-Kane}
\beq
\label{eq:53}
\Gamma^1 = \tau^x,\,\, \Gamma^2 = \tau^y,\,\, \Gamma^3 = \tau^z \sigma^x,\,\, \Gamma^4 = \tau^z \sigma^y,\,\,
\Gamma^5 = \tau^z \sigma^z, 
\eeq
where $\btau$ and $\bsigma$ are Pauli matrices, acting on the orbital and spin degrees of freedom correspondingly and 
the parity operator $P = \tau^x$. 
The multilayer Hamiltonian Eq.~\eqref{eq:1}, without the time-reversal breaking term $b \sigma^z$, is precisely of this form with 
\beqa
\label{eq:54} 
d_0(\bk)&=&0, \nonumber \\
d_1(\bk)&=&t_S + t_D \cos(k_z d), \nonumber \\
d_2(\bk)&=&- t_D \sin(k_z d), \nonumber \\
d_3(\bk)&=&v_F k_y, \nonumber \\
d_4(\bk)&=&-v_F k_x, \nonumber \\
d_5(\bk)&=&0. 
\eeqa

Suppose a Dirac band touching point is realized at a TRIM $\bGamma$. 
In this case we have $d_{a}(\bGamma) = 0$ (we set $d_0(\bGamma) = 0$, which simply defines the zero of energy). 
The particle-hole symmetry violating term $d_0(\bk)$ and the Dirac mass term $d_1(\bk)$ are both even under parity. 
Their Taylor expansion in the vicinity of $\bGamma$ thus starts
with terms, quadratic in $ \delta \bk = \bk - \bGamma$, which may, to the first approximation, be ignored. 
The other four coefficients are all odd under parity and their Taylor expansions near $\bGamma$ must thus 
involve terms, linear in $\delta \bk$, or cubic and higher. 
The matrix coefficients of the linear terms, proportional to $\delta k_{x,y,z}$, determine the three Dirac $\gamma$-matrices 
$\gamma^i$ with $i=1,2,3$ as
\beq
\label{eq:55}
H(\bk) \approx \gamma^0( v_x \gamma^1 \delta k_x + v_y \gamma^2 \delta k_y + v_z \gamma^3 \delta k_z), 
\eeq
where $\gamma^0 \equiv \Gamma^1$, and $v_{x,y,z}$ are the Fermi velocities, corresponding to the $x,y,z$ directions.
Once the four basic Dirac $\gamma$-matrices are defined, the chiral charge operator for this particular Dirac point is then given by
\beq
\label{eq:56}
\gamma^5 = i \gamma^0 \gamma^1 \gamma^2 \gamma^3. 
\eeq
Since $\gamma^5$ anticommutes with the four Dirac matrices $\gamma^{\mu}$, it {\em commutes} with $H(\bk)$, but only as 
long as the terms, quadratic in $\delta \bk$, may be ignored. 
Once the quadratic (and higher-order) terms are included, the Dirac mass term, proportional to $\gamma^0$, is present 
in $H(\bk)$ and $[H(\bk), \gamma^5] \neq 0$, i.e. the chiral charge is not conserved. 
It is in this sense that any Dirac point at TRIM has an emergent chiral symmetry. 
If the Dirac point is split into a pair of Weyl points, the resulting Weyl semimetal will also possess an approximate 
chiral symmetry, as long as the splitting is small compared to the size of the BZ, and thus the chiral symmetry violating 
terms in the Taylor expansion of $d_1(\bk)$ may still be ignored. 

In this paper we have studied the simplest case of only a single pair of Weyl points. 
However, our arguments may be easily generalized to multiple pairs, each emerging from its own parent Dirac point at a 
specific TRIM. Since BCS superconductivity in our context involves pairing between parity-related eigenstates and TRIM are parity-invariant, the BCS pairing term does not mix states near different Dirac points. 
Thus the problem reduces to the one considered above. 

In conclusion, we have studied the electromagnetic response of a topological Weyl superconductor with broken time reversal 
symmetry. We have found that, under natural conditions, the anomalous Hall conductivity of a Weyl superconductor 
coincides with that of a nonsuperconducting Weyl semimetal. 
We have connected this unexpected result with the appearance of a new conserved quantity in a Weyl semimetal, the 
chiral charge. 

\section{Acknowledgments}
Financial support was provided by NSERC of Canada (GB and AAB) and the Swiss SNF and the NCCR Quantum Science and 
Technology (AAZ). 
\bibliography{references}

\begin{thebibliography}{50}%
\makeatletter
\providecommand \@ifxundefined [1]{%
 \@ifx{#1\undefined}
}%
\providecommand \@ifnum [1]{%
 \ifnum #1\expandafter \@firstoftwo
 \else \expandafter \@secondoftwo
 \fi
}%
\providecommand \@ifx [1]{%
 \ifx #1\expandafter \@firstoftwo
 \else \expandafter \@secondoftwo
 \fi
}%
\providecommand \natexlab [1]{#1}%
\providecommand \enquote  [1]{``#1''}%
\providecommand \bibnamefont  [1]{#1}%
\providecommand \bibfnamefont [1]{#1}%
\providecommand \citenamefont [1]{#1}%
\providecommand \href@noop [0]{\@secondoftwo}%
\providecommand \href [0]{\begingroup \@sanitize@url \@href}%
\providecommand \@href[1]{\@@startlink{#1}\@@href}%
\providecommand \@@href[1]{\endgroup#1\@@endlink}%
\providecommand \@sanitize@url [0]{\catcode `\\12\catcode `\$12\catcode
  `\&12\catcode `\#12\catcode `\^12\catcode `\_12\catcode `\%12\relax}%
\providecommand \@@startlink[1]{}%
\providecommand \@@endlink[0]{}%
\providecommand \url  [0]{\begingroup\@sanitize@url \@url }%
\providecommand \@url [1]{\endgroup\@href {#1}{\urlprefix }}%
\providecommand \urlprefix  [0]{URL }%
\providecommand \Eprint [0]{\href }%
\providecommand \doibase [0]{http://dx.doi.org/}%
\providecommand \selectlanguage [0]{\@gobble}%
\providecommand \bibinfo  [0]{\@secondoftwo}%
\providecommand \bibfield  [0]{\@secondoftwo}%
\providecommand \translation [1]{[#1]}%
\providecommand \BibitemOpen [0]{}%
\providecommand \bibitemStop [0]{}%
\providecommand \bibitemNoStop [0]{.\EOS\space}%
\providecommand \EOS [0]{\spacefactor3000\relax}%
\providecommand \BibitemShut  [1]{\csname bibitem#1\endcsname}%
\let\auto@bib@innerbib\@empty
%</preamble>
\bibitem [{\citenamefont {Hasan}\ and\ \citenamefont {Kane}(2010)}]{Hasan10}%
  \BibitemOpen
  \bibfield  {author} {\bibinfo {author} {\bibfnamefont {M.~Z.}\ \bibnamefont
  {Hasan}}\ and\ \bibinfo {author} {\bibfnamefont {C.~L.}\ \bibnamefont
  {Kane}},\ }\href {\doibase 10.1103/RevModPhys.82.3045} {\bibfield  {journal}
  {\bibinfo  {journal} {Rev. Mod. Phys.}\ }\textbf {\bibinfo {volume} {82}},\
  \bibinfo {pages} {3045} (\bibinfo {year} {2010})}\BibitemShut {NoStop}%
\bibitem [{\citenamefont {Qi}\ and\ \citenamefont {Zhang}(2011)}]{Qi11}%
  \BibitemOpen
  \bibfield  {author} {\bibinfo {author} {\bibfnamefont {X.-L.}\ \bibnamefont
  {Qi}}\ and\ \bibinfo {author} {\bibfnamefont {S.-C.}\ \bibnamefont {Zhang}},\
  }\href {\doibase 10.1103/RevModPhys.83.1057} {\bibfield  {journal} {\bibinfo
  {journal} {Rev. Mod. Phys.}\ }\textbf {\bibinfo {volume} {83}},\ \bibinfo
  {pages} {1057} (\bibinfo {year} {2011})}\BibitemShut {NoStop}%
\bibitem [{\citenamefont {Volovik}(1988)}]{Volovik88}%
  \BibitemOpen
  \bibfield  {author} {\bibinfo {author} {\bibfnamefont {G.~E.}\ \bibnamefont
  {Volovik}},\ }\href@noop {} {\bibfield  {journal} {\bibinfo  {journal} {Sov.
  Phys. JETP}\ }\textbf {\bibinfo {volume} {67}},\ \bibinfo {pages} {1804}
  (\bibinfo {year} {1988})}\BibitemShut {NoStop}%
\bibitem [{\citenamefont {Volovik}(2003)}]{Volovik03}%
  \BibitemOpen
  \bibfield  {author} {\bibinfo {author} {\bibfnamefont {G.}~\bibnamefont
  {Volovik}},\ }\href@noop {} {\emph {\bibinfo {title} {The Universe in a
  Helium Droplet}}}\ (\bibinfo  {publisher} {Oxford: Clarendon},\ \bibinfo
  {year} {2003})\BibitemShut {NoStop}%
\bibitem [{\citenamefont {Haldane}(2004)}]{Haldane04}%
  \BibitemOpen
  \bibfield  {author} {\bibinfo {author} {\bibfnamefont {F.~D.~M.}\
  \bibnamefont {Haldane}},\ }\href {\doibase 10.1103/PhysRevLett.93.206602}
  {\bibfield  {journal} {\bibinfo  {journal} {Phys. Rev. Lett.}\ }\textbf
  {\bibinfo {volume} {93}},\ \bibinfo {pages} {206602} (\bibinfo {year}
  {2004})}\BibitemShut {NoStop}%
\bibitem [{\citenamefont {Volovik}(2007)}]{Volovik07}%
  \BibitemOpen
  \bibfield  {author} {\bibinfo {author} {\bibfnamefont {G.~E.}\ \bibnamefont
  {Volovik}},\ }in\ \href {\doibase 10.1007/3-540-70859-6_3} {\emph {\bibinfo
  {booktitle} {Quantum Analogues: From Phase Transitions to Black Holes and
  Cosmology}}},\ \bibinfo {series} {Lecture Notes in Physics}, Vol.\ \bibinfo
  {volume} {718},\ \bibinfo {editor} {edited by\ \bibinfo {editor}
  {\bibfnamefont {W.}~\bibnamefont {Unruh}}\ and\ \bibinfo {editor}
  {\bibfnamefont {R.}~\bibnamefont {Schützhold}}}\ (\bibinfo  {publisher}
  {Springer Berlin Heidelberg},\ \bibinfo {year} {2007})\BibitemShut {NoStop}%
\bibitem [{\citenamefont {Murakami}(2007)}]{Murakami07}%
  \BibitemOpen
  \bibfield  {author} {\bibinfo {author} {\bibfnamefont {S.}~\bibnamefont
  {Murakami}},\ }\href {http://stacks.iop.org/1367-2630/9/i=9/a=356} {\bibfield
   {journal} {\bibinfo  {journal} {New Journal of Physics}\ }\textbf {\bibinfo
  {volume} {9}},\ \bibinfo {pages} {356} (\bibinfo {year} {2007})}\BibitemShut
  {NoStop}%
\bibitem [{\citenamefont {Wan}\ \emph {et~al.}(2011)\citenamefont {Wan},
  \citenamefont {Turner}, \citenamefont {Vishwanath},\ and\ \citenamefont
  {Savrasov}}]{Wan11}%
  \BibitemOpen
  \bibfield  {author} {\bibinfo {author} {\bibfnamefont {X.}~\bibnamefont
  {Wan}}, \bibinfo {author} {\bibfnamefont {A.~M.}\ \bibnamefont {Turner}},
  \bibinfo {author} {\bibfnamefont {A.}~\bibnamefont {Vishwanath}}, \ and\
  \bibinfo {author} {\bibfnamefont {S.~Y.}\ \bibnamefont {Savrasov}},\ }\href
  {\doibase 10.1103/PhysRevB.83.205101} {\bibfield  {journal} {\bibinfo
  {journal} {Phys. Rev. B}\ }\textbf {\bibinfo {volume} {83}},\ \bibinfo
  {pages} {205101} (\bibinfo {year} {2011})}\BibitemShut {NoStop}%
\bibitem [{\citenamefont {Yang}\ \emph {et~al.}(2011)\citenamefont {Yang},
  \citenamefont {Lu},\ and\ \citenamefont {Ran}}]{Ran11}%
  \BibitemOpen
  \bibfield  {author} {\bibinfo {author} {\bibfnamefont {K.-Y.}\ \bibnamefont
  {Yang}}, \bibinfo {author} {\bibfnamefont {Y.-M.}\ \bibnamefont {Lu}}, \ and\
  \bibinfo {author} {\bibfnamefont {Y.}~\bibnamefont {Ran}},\ }\href {\doibase
  10.1103/PhysRevB.84.075129} {\bibfield  {journal} {\bibinfo  {journal} {Phys.
  Rev. B}\ }\textbf {\bibinfo {volume} {84}},\ \bibinfo {pages} {075129}
  (\bibinfo {year} {2011})}\BibitemShut {NoStop}%
\bibitem [{\citenamefont {Burkov}\ and\ \citenamefont
  {Balents}(2011)}]{Burkov11-1}%
  \BibitemOpen
  \bibfield  {author} {\bibinfo {author} {\bibfnamefont {A.~A.}\ \bibnamefont
  {Burkov}}\ and\ \bibinfo {author} {\bibfnamefont {L.}~\bibnamefont
  {Balents}},\ }\href {\doibase 10.1103/PhysRevLett.107.127205} {\bibfield
  {journal} {\bibinfo  {journal} {Phys. Rev. Lett.}\ }\textbf {\bibinfo
  {volume} {107}},\ \bibinfo {pages} {127205} (\bibinfo {year}
  {2011})}\BibitemShut {NoStop}%
\bibitem [{\citenamefont {Burkov}\ \emph {et~al.}(2011)\citenamefont {Burkov},
  \citenamefont {Hook},\ and\ \citenamefont {Balents}}]{Burkov11-2}%
  \BibitemOpen
  \bibfield  {author} {\bibinfo {author} {\bibfnamefont {A.~A.}\ \bibnamefont
  {Burkov}}, \bibinfo {author} {\bibfnamefont {M.~D.}\ \bibnamefont {Hook}}, \
  and\ \bibinfo {author} {\bibfnamefont {L.}~\bibnamefont {Balents}},\ }\href
  {\doibase 10.1103/PhysRevB.84.235126} {\bibfield  {journal} {\bibinfo
  {journal} {Phys. Rev. B}\ }\textbf {\bibinfo {volume} {84}},\ \bibinfo
  {pages} {235126} (\bibinfo {year} {2011})}\BibitemShut {NoStop}%
\bibitem [{\citenamefont {Xu}\ \emph {et~al.}(2011)\citenamefont {Xu},
  \citenamefont {Weng}, \citenamefont {Wang}, \citenamefont {Dai},\ and\
  \citenamefont {Fang}}]{Xu11}%
  \BibitemOpen
  \bibfield  {author} {\bibinfo {author} {\bibfnamefont {G.}~\bibnamefont
  {Xu}}, \bibinfo {author} {\bibfnamefont {H.}~\bibnamefont {Weng}}, \bibinfo
  {author} {\bibfnamefont {Z.}~\bibnamefont {Wang}}, \bibinfo {author}
  {\bibfnamefont {X.}~\bibnamefont {Dai}}, \ and\ \bibinfo {author}
  {\bibfnamefont {Z.}~\bibnamefont {Fang}},\ }\href {\doibase
  10.1103/PhysRevLett.107.186806} {\bibfield  {journal} {\bibinfo  {journal}
  {Phys. Rev. Lett.}\ }\textbf {\bibinfo {volume} {107}},\ \bibinfo {pages}
  {186806} (\bibinfo {year} {2011})}\BibitemShut {NoStop}%
\bibitem [{\citenamefont {Young}\ \emph {et~al.}(2012)\citenamefont {Young},
  \citenamefont {Zaheer}, \citenamefont {Teo}, \citenamefont {Kane},
  \citenamefont {Mele},\ and\ \citenamefont {Rappe}}]{Kane12}%
  \BibitemOpen
  \bibfield  {author} {\bibinfo {author} {\bibfnamefont {S.~M.}\ \bibnamefont
  {Young}}, \bibinfo {author} {\bibfnamefont {S.}~\bibnamefont {Zaheer}},
  \bibinfo {author} {\bibfnamefont {J.~C.~Y.}\ \bibnamefont {Teo}}, \bibinfo
  {author} {\bibfnamefont {C.~L.}\ \bibnamefont {Kane}}, \bibinfo {author}
  {\bibfnamefont {E.~J.}\ \bibnamefont {Mele}}, \ and\ \bibinfo {author}
  {\bibfnamefont {A.~M.}\ \bibnamefont {Rappe}},\ }\href {\doibase
  10.1103/PhysRevLett.108.140405} {\bibfield  {journal} {\bibinfo  {journal}
  {Phys. Rev. Lett.}\ }\textbf {\bibinfo {volume} {108}},\ \bibinfo {pages}
  {140405} (\bibinfo {year} {2012})}\BibitemShut {NoStop}%
\bibitem [{\citenamefont {Wang}\ \emph {et~al.}(2012)\citenamefont {Wang},
  \citenamefont {Sun}, \citenamefont {Chen}, \citenamefont {Franchini},
  \citenamefont {Xu}, \citenamefont {Weng}, \citenamefont {Dai},\ and\
  \citenamefont {Fang}}]{Fang12}%
  \BibitemOpen
  \bibfield  {author} {\bibinfo {author} {\bibfnamefont {Z.}~\bibnamefont
  {Wang}}, \bibinfo {author} {\bibfnamefont {Y.}~\bibnamefont {Sun}}, \bibinfo
  {author} {\bibfnamefont {X.-Q.}\ \bibnamefont {Chen}}, \bibinfo {author}
  {\bibfnamefont {C.}~\bibnamefont {Franchini}}, \bibinfo {author}
  {\bibfnamefont {G.}~\bibnamefont {Xu}}, \bibinfo {author} {\bibfnamefont
  {H.}~\bibnamefont {Weng}}, \bibinfo {author} {\bibfnamefont {X.}~\bibnamefont
  {Dai}}, \ and\ \bibinfo {author} {\bibfnamefont {Z.}~\bibnamefont {Fang}},\
  }\href {\doibase 10.1103/PhysRevB.85.195320} {\bibfield  {journal} {\bibinfo
  {journal} {Phys. Rev. B}\ }\textbf {\bibinfo {volume} {85}},\ \bibinfo
  {pages} {195320} (\bibinfo {year} {2012})}\BibitemShut {NoStop}%
\bibitem [{\citenamefont {Wang}\ \emph {et~al.}(2013)\citenamefont {Wang},
  \citenamefont {Weng}, \citenamefont {Wu}, \citenamefont {Dai},\ and\
  \citenamefont {Fang}}]{Fang13}%
  \BibitemOpen
  \bibfield  {author} {\bibinfo {author} {\bibfnamefont {Z.}~\bibnamefont
  {Wang}}, \bibinfo {author} {\bibfnamefont {H.}~\bibnamefont {Weng}}, \bibinfo
  {author} {\bibfnamefont {Q.}~\bibnamefont {Wu}}, \bibinfo {author}
  {\bibfnamefont {X.}~\bibnamefont {Dai}}, \ and\ \bibinfo {author}
  {\bibfnamefont {Z.}~\bibnamefont {Fang}},\ }\href {\doibase
  10.1103/PhysRevB.88.125427} {\bibfield  {journal} {\bibinfo  {journal} {Phys.
  Rev. B}\ }\textbf {\bibinfo {volume} {88}},\ \bibinfo {pages} {125427}
  (\bibinfo {year} {2013})}\BibitemShut {NoStop}%
\bibitem [{\citenamefont {Xu}\ \emph {et~al.}(2015{\natexlab{a}})\citenamefont
  {Xu}, \citenamefont {Belopolski}, \citenamefont {Alidoust}, \citenamefont
  {Neupane}, \citenamefont {Bian}, \citenamefont {Zhang}, \citenamefont
  {Sankar}, \citenamefont {Chang}, \citenamefont {Yuan}, \citenamefont {Lee},
  \citenamefont {Huang}, \citenamefont {Zheng}, \citenamefont {Ma},
  \citenamefont {Sanchez}, \citenamefont {Wang}, \citenamefont {Bansil},
  \citenamefont {Chou}, \citenamefont {Shibayev}, \citenamefont {Lin},
  \citenamefont {Jia},\ and\ \citenamefont {Hasan}}]{HasanTaAs}%
  \BibitemOpen
  \bibfield  {author} {\bibinfo {author} {\bibfnamefont {S.-Y.}\ \bibnamefont
  {Xu}}, \bibinfo {author} {\bibfnamefont {I.}~\bibnamefont {Belopolski}},
  \bibinfo {author} {\bibfnamefont {N.}~\bibnamefont {Alidoust}}, \bibinfo
  {author} {\bibfnamefont {M.}~\bibnamefont {Neupane}}, \bibinfo {author}
  {\bibfnamefont {G.}~\bibnamefont {Bian}}, \bibinfo {author} {\bibfnamefont
  {C.}~\bibnamefont {Zhang}}, \bibinfo {author} {\bibfnamefont
  {R.}~\bibnamefont {Sankar}}, \bibinfo {author} {\bibfnamefont
  {G.}~\bibnamefont {Chang}}, \bibinfo {author} {\bibfnamefont
  {Z.}~\bibnamefont {Yuan}}, \bibinfo {author} {\bibfnamefont {C.-C.}\
  \bibnamefont {Lee}}, \bibinfo {author} {\bibfnamefont {S.-M.}\ \bibnamefont
  {Huang}}, \bibinfo {author} {\bibfnamefont {H.}~\bibnamefont {Zheng}},
  \bibinfo {author} {\bibfnamefont {J.}~\bibnamefont {Ma}}, \bibinfo {author}
  {\bibfnamefont {D.~S.}\ \bibnamefont {Sanchez}}, \bibinfo {author}
  {\bibfnamefont {B.}~\bibnamefont {Wang}}, \bibinfo {author} {\bibfnamefont
  {A.}~\bibnamefont {Bansil}}, \bibinfo {author} {\bibfnamefont
  {F.}~\bibnamefont {Chou}}, \bibinfo {author} {\bibfnamefont {P.~P.}\
  \bibnamefont {Shibayev}}, \bibinfo {author} {\bibfnamefont {H.}~\bibnamefont
  {Lin}}, \bibinfo {author} {\bibfnamefont {S.}~\bibnamefont {Jia}}, \ and\
  \bibinfo {author} {\bibfnamefont {M.~Z.}\ \bibnamefont {Hasan}},\ }\href
  {\doibase 10.1126/science.aaa9297} {\bibfield  {journal} {\bibinfo  {journal}
  {Science}\ }\textbf {\bibinfo {volume} {349}},\ \bibinfo {pages} {613}
  (\bibinfo {year} {2015}{\natexlab{a}})}\BibitemShut {NoStop}%
\bibitem [{\citenamefont {Neupane}\ \emph {et~al.}(2014)\citenamefont
  {Neupane}, \citenamefont {Xu}, \citenamefont {Sankar}, \citenamefont
  {Alidoust}, \citenamefont {Bian}, \citenamefont {Liu}, \citenamefont
  {Belopolski}, \citenamefont {Chang}, \citenamefont {Jeng}, \citenamefont
  {Lin}, \citenamefont {Bansil}, \citenamefont {Chou},\ and\ \citenamefont
  {Hasan}}]{Neupane14}%
  \BibitemOpen
  \bibfield  {author} {\bibinfo {author} {\bibfnamefont {M.}~\bibnamefont
  {Neupane}}, \bibinfo {author} {\bibfnamefont {S.-Y.}\ \bibnamefont {Xu}},
  \bibinfo {author} {\bibfnamefont {R.}~\bibnamefont {Sankar}}, \bibinfo
  {author} {\bibfnamefont {N.}~\bibnamefont {Alidoust}}, \bibinfo {author}
  {\bibfnamefont {G.}~\bibnamefont {Bian}}, \bibinfo {author} {\bibfnamefont
  {C.}~\bibnamefont {Liu}}, \bibinfo {author} {\bibfnamefont {I.}~\bibnamefont
  {Belopolski}}, \bibinfo {author} {\bibfnamefont {T.-R.}\ \bibnamefont
  {Chang}}, \bibinfo {author} {\bibfnamefont {H.-T.}\ \bibnamefont {Jeng}},
  \bibinfo {author} {\bibfnamefont {H.}~\bibnamefont {Lin}}, \bibinfo {author}
  {\bibfnamefont {A.}~\bibnamefont {Bansil}}, \bibinfo {author} {\bibfnamefont
  {F.}~\bibnamefont {Chou}}, \ and\ \bibinfo {author} {\bibfnamefont {M.~Z.}\
  \bibnamefont {Hasan}},\ }\href {http://dx.doi.org/10.1038/ncomms4786}
  {\bibfield  {journal} {\bibinfo  {journal} {Nat. Commun.}\ }\textbf {\bibinfo
  {volume} {5}} (\bibinfo {year} {2014})}\BibitemShut {NoStop}%
\bibitem [{\citenamefont {Lv}\ \emph {et~al.}(2015{\natexlab{a}})\citenamefont
  {Lv}, \citenamefont {Xu}, \citenamefont {Weng}, \citenamefont {Ma},
  \citenamefont {Richard}, \citenamefont {Huang}, \citenamefont {Zhao},
  \citenamefont {Chen}, \citenamefont {Matt}, \citenamefont {Bisti},
  \citenamefont {Strocov}, \citenamefont {Mesot}, \citenamefont {Fang},
  \citenamefont {Dai}, \citenamefont {Qian}, \citenamefont {Shi},\ and\
  \citenamefont {Ding}}]{DingTaAs2}%
  \BibitemOpen
  \bibfield  {author} {\bibinfo {author} {\bibfnamefont {B.~Q.}\ \bibnamefont
  {Lv}}, \bibinfo {author} {\bibfnamefont {N.}~\bibnamefont {Xu}}, \bibinfo
  {author} {\bibfnamefont {H.~M.}\ \bibnamefont {Weng}}, \bibinfo {author}
  {\bibfnamefont {J.~Z.}\ \bibnamefont {Ma}}, \bibinfo {author} {\bibfnamefont
  {P.}~\bibnamefont {Richard}}, \bibinfo {author} {\bibfnamefont {X.~C.}\
  \bibnamefont {Huang}}, \bibinfo {author} {\bibfnamefont {L.~X.}\ \bibnamefont
  {Zhao}}, \bibinfo {author} {\bibfnamefont {G.~F.}\ \bibnamefont {Chen}},
  \bibinfo {author} {\bibfnamefont {C.~E.}\ \bibnamefont {Matt}}, \bibinfo
  {author} {\bibfnamefont {F.}~\bibnamefont {Bisti}}, \bibinfo {author}
  {\bibfnamefont {V.~N.}\ \bibnamefont {Strocov}}, \bibinfo {author}
  {\bibfnamefont {J.}~\bibnamefont {Mesot}}, \bibinfo {author} {\bibfnamefont
  {Z.}~\bibnamefont {Fang}}, \bibinfo {author} {\bibfnamefont {X.}~\bibnamefont
  {Dai}}, \bibinfo {author} {\bibfnamefont {T.}~\bibnamefont {Qian}}, \bibinfo
  {author} {\bibfnamefont {M.}~\bibnamefont {Shi}}, \ and\ \bibinfo {author}
  {\bibfnamefont {H.}~\bibnamefont {Ding}},\ }\href
  {http://dx.doi.org/10.1038/nphys3426} {\bibfield  {journal} {\bibinfo
  {journal} {Nat Phys}\ }\textbf {\bibinfo {volume} {11}},\ \bibinfo {pages}
  {724} (\bibinfo {year} {2015}{\natexlab{a}})}\BibitemShut {NoStop}%
\bibitem [{\citenamefont {Lv}\ \emph {et~al.}(2015{\natexlab{b}})\citenamefont
  {Lv}, \citenamefont {Weng}, \citenamefont {Fu}, \citenamefont {Wang},
  \citenamefont {Miao}, \citenamefont {Ma}, \citenamefont {Richard},
  \citenamefont {Huang}, \citenamefont {Zhao}, \citenamefont {Chen},
  \citenamefont {Fang}, \citenamefont {Dai}, \citenamefont {Qian},\ and\
  \citenamefont {Ding}}]{DingTaAs}%
  \BibitemOpen
  \bibfield  {author} {\bibinfo {author} {\bibfnamefont {B.~Q.}\ \bibnamefont
  {Lv}}, \bibinfo {author} {\bibfnamefont {H.~M.}\ \bibnamefont {Weng}},
  \bibinfo {author} {\bibfnamefont {B.~B.}\ \bibnamefont {Fu}}, \bibinfo
  {author} {\bibfnamefont {X.~P.}\ \bibnamefont {Wang}}, \bibinfo {author}
  {\bibfnamefont {H.}~\bibnamefont {Miao}}, \bibinfo {author} {\bibfnamefont
  {J.}~\bibnamefont {Ma}}, \bibinfo {author} {\bibfnamefont {P.}~\bibnamefont
  {Richard}}, \bibinfo {author} {\bibfnamefont {X.~C.}\ \bibnamefont {Huang}},
  \bibinfo {author} {\bibfnamefont {L.~X.}\ \bibnamefont {Zhao}}, \bibinfo
  {author} {\bibfnamefont {G.~F.}\ \bibnamefont {Chen}}, \bibinfo {author}
  {\bibfnamefont {Z.}~\bibnamefont {Fang}}, \bibinfo {author} {\bibfnamefont
  {X.}~\bibnamefont {Dai}}, \bibinfo {author} {\bibfnamefont {T.}~\bibnamefont
  {Qian}}, \ and\ \bibinfo {author} {\bibfnamefont {H.}~\bibnamefont {Ding}},\
  }\href {\doibase 10.1103/PhysRevX.5.031013} {\bibfield  {journal} {\bibinfo
  {journal} {Phys. Rev. X}\ }\textbf {\bibinfo {volume} {5}},\ \bibinfo {pages}
  {031013} (\bibinfo {year} {2015}{\natexlab{b}})}\BibitemShut {NoStop}%
\bibitem [{\citenamefont {Lu}\ \emph {et~al.}(2015{\natexlab{a}})\citenamefont
  {Lu}, \citenamefont {Wang}, \citenamefont {Ye}, \citenamefont {Ran},
  \citenamefont {Fu}, \citenamefont {Joannopoulos},\ and\ \citenamefont
  {Solja{\v c}i{\'c}}}]{Lu15}%
  \BibitemOpen
  \bibfield  {author} {\bibinfo {author} {\bibfnamefont {L.}~\bibnamefont
  {Lu}}, \bibinfo {author} {\bibfnamefont {Z.}~\bibnamefont {Wang}}, \bibinfo
  {author} {\bibfnamefont {D.}~\bibnamefont {Ye}}, \bibinfo {author}
  {\bibfnamefont {L.}~\bibnamefont {Ran}}, \bibinfo {author} {\bibfnamefont
  {L.}~\bibnamefont {Fu}}, \bibinfo {author} {\bibfnamefont {J.~D.}\
  \bibnamefont {Joannopoulos}}, \ and\ \bibinfo {author} {\bibfnamefont
  {M.}~\bibnamefont {Solja{\v c}i{\'c}}},\ }\href {\doibase
  10.1126/science.aaa9273} {\bibfield  {journal} {\bibinfo  {journal}
  {Science}\ }\textbf {\bibinfo {volume} {349}},\ \bibinfo {pages} {622}
  (\bibinfo {year} {2015}{\natexlab{a}})}\BibitemShut {NoStop}%
\bibitem [{\citenamefont {{Haldane}}(2014)}]{Haldane14}%
  \BibitemOpen
  \bibfield  {author} {\bibinfo {author} {\bibfnamefont {F.~D.~M.}\
  \bibnamefont {{Haldane}}},\ }\href@noop {} {\bibfield  {journal} {\bibinfo
  {journal} {ArXiv e-prints}\ } (\bibinfo {year} {2014})},\ \Eprint
  {http://arxiv.org/abs/1401.0529} {arXiv:1401.0529 [cond-mat.str-el]}
  \BibitemShut {NoStop}%
\bibitem [{\citenamefont {Chen}\ \emph {et~al.}(2013)\citenamefont {Chen},
  \citenamefont {Bergman},\ and\ \citenamefont {Burkov}}]{Burkov13}%
  \BibitemOpen
  \bibfield  {author} {\bibinfo {author} {\bibfnamefont {Y.}~\bibnamefont
  {Chen}}, \bibinfo {author} {\bibfnamefont {D.~L.}\ \bibnamefont {Bergman}}, \
  and\ \bibinfo {author} {\bibfnamefont {A.~A.}\ \bibnamefont {Burkov}},\
  }\href {\doibase 10.1103/PhysRevB.88.125110} {\bibfield  {journal} {\bibinfo
  {journal} {Phys. Rev. B}\ }\textbf {\bibinfo {volume} {88}},\ \bibinfo
  {pages} {125110} (\bibinfo {year} {2013})}\BibitemShut {NoStop}%
\bibitem [{\citenamefont {Burkov}(2014{\natexlab{a}})}]{Burkov14-1}%
  \BibitemOpen
  \bibfield  {author} {\bibinfo {author} {\bibfnamefont {A.~A.}\ \bibnamefont
  {Burkov}},\ }\href {\doibase 10.1103/PhysRevB.89.155104} {\bibfield
  {journal} {\bibinfo  {journal} {Phys. Rev. B}\ }\textbf {\bibinfo {volume}
  {89}},\ \bibinfo {pages} {155104} (\bibinfo {year}
  {2014}{\natexlab{a}})}\BibitemShut {NoStop}%
\bibitem [{\citenamefont {Burkov}(2014{\natexlab{b}})}]{Burkov14-2}%
  \BibitemOpen
  \bibfield  {author} {\bibinfo {author} {\bibfnamefont {A.~A.}\ \bibnamefont
  {Burkov}},\ }\href {\doibase 10.1103/PhysRevLett.113.187202} {\bibfield
  {journal} {\bibinfo  {journal} {Phys. Rev. Lett.}\ }\textbf {\bibinfo
  {volume} {113}},\ \bibinfo {pages} {187202} (\bibinfo {year}
  {2014}{\natexlab{b}})}\BibitemShut {NoStop}%
\bibitem [{\citenamefont {{Shekhar}}\ \emph {et~al.}(2016)\citenamefont
  {{Shekhar}}, \citenamefont {{Nayak}}, \citenamefont {{Singh}}, \citenamefont
  {{Kumar}}, \citenamefont {{Wu}}, \citenamefont {{Zhang}}, \citenamefont
  {{Komarek}}, \citenamefont {{Kampert}}, \citenamefont {{Skourski}},
  \citenamefont {{Wosnitza}}, \citenamefont {{Schnelle}}, \citenamefont
  {{McCollam}}, \citenamefont {{Zeitler}}, \citenamefont {{Kubler}},
  \citenamefont {{Parkin}}, \citenamefont {{Yan}},\ and\ \citenamefont
  {{Felser}}}]{Felser16}%
  \BibitemOpen
  \bibfield  {author} {\bibinfo {author} {\bibfnamefont {C.}~\bibnamefont
  {{Shekhar}}}, \bibinfo {author} {\bibfnamefont {A.~K.}\ \bibnamefont
  {{Nayak}}}, \bibinfo {author} {\bibfnamefont {S.}~\bibnamefont {{Singh}}},
  \bibinfo {author} {\bibfnamefont {N.}~\bibnamefont {{Kumar}}}, \bibinfo
  {author} {\bibfnamefont {S.-C.}\ \bibnamefont {{Wu}}}, \bibinfo {author}
  {\bibfnamefont {Y.}~\bibnamefont {{Zhang}}}, \bibinfo {author} {\bibfnamefont
  {A.~C.}\ \bibnamefont {{Komarek}}}, \bibinfo {author} {\bibfnamefont
  {E.}~\bibnamefont {{Kampert}}}, \bibinfo {author} {\bibfnamefont
  {Y.}~\bibnamefont {{Skourski}}}, \bibinfo {author} {\bibfnamefont
  {J.}~\bibnamefont {{Wosnitza}}}, \bibinfo {author} {\bibfnamefont
  {W.}~\bibnamefont {{Schnelle}}}, \bibinfo {author} {\bibfnamefont
  {A.}~\bibnamefont {{McCollam}}}, \bibinfo {author} {\bibfnamefont
  {U.}~\bibnamefont {{Zeitler}}}, \bibinfo {author} {\bibfnamefont
  {J.}~\bibnamefont {{Kubler}}}, \bibinfo {author} {\bibfnamefont {S.~S.~P.}\
  \bibnamefont {{Parkin}}}, \bibinfo {author} {\bibfnamefont {B.}~\bibnamefont
  {{Yan}}}, \ and\ \bibinfo {author} {\bibfnamefont {C.}~\bibnamefont
  {{Felser}}},\ }\href@noop {} {\bibfield  {journal} {\bibinfo  {journal}
  {ArXiv e-prints}\ } (\bibinfo {year} {2016})},\ \Eprint
  {http://arxiv.org/abs/1604.01641} {arXiv:1604.01641 [cond-mat.mtrl-sci]}
  \BibitemShut {NoStop}%
\bibitem [{\citenamefont {Son}\ and\ \citenamefont {Spivak}(2013)}]{Spivak12}%
  \BibitemOpen
  \bibfield  {author} {\bibinfo {author} {\bibfnamefont {D.~T.}\ \bibnamefont
  {Son}}\ and\ \bibinfo {author} {\bibfnamefont {B.~Z.}\ \bibnamefont
  {Spivak}},\ }\href {\doibase 10.1103/PhysRevB.88.104412} {\bibfield
  {journal} {\bibinfo  {journal} {Phys. Rev. B}\ }\textbf {\bibinfo {volume}
  {88}},\ \bibinfo {pages} {104412} (\bibinfo {year} {2013})}\BibitemShut
  {NoStop}%
\bibitem [{\citenamefont {Burkov}(2014{\natexlab{c}})}]{Burkov_lmr_prl}%
  \BibitemOpen
  \bibfield  {author} {\bibinfo {author} {\bibfnamefont {A.~A.}\ \bibnamefont
  {Burkov}},\ }\href {\doibase 10.1103/PhysRevLett.113.247203} {\bibfield
  {journal} {\bibinfo  {journal} {Phys. Rev. Lett.}\ }\textbf {\bibinfo
  {volume} {113}},\ \bibinfo {pages} {247203} (\bibinfo {year}
  {2014}{\natexlab{c}})}\BibitemShut {NoStop}%
\bibitem [{\citenamefont {Burkov}(2015)}]{Burkov_lmr_prb}%
  \BibitemOpen
  \bibfield  {author} {\bibinfo {author} {\bibfnamefont {A.~A.}\ \bibnamefont
  {Burkov}},\ }\href {\doibase 10.1103/PhysRevB.91.245157} {\bibfield
  {journal} {\bibinfo  {journal} {Phys. Rev. B}\ }\textbf {\bibinfo {volume}
  {91}},\ \bibinfo {pages} {245157} (\bibinfo {year} {2015})}\BibitemShut
  {NoStop}%
\bibitem [{\citenamefont {Xiong}\ \emph {et~al.}(2015)\citenamefont {Xiong},
  \citenamefont {Kushwaha}, \citenamefont {Liang}, \citenamefont {Krizan},
  \citenamefont {Hirschberger}, \citenamefont {Wang}, \citenamefont {Cava},\
  and\ \citenamefont {Ong}}]{Ong_anomaly}%
  \BibitemOpen
  \bibfield  {author} {\bibinfo {author} {\bibfnamefont {J.}~\bibnamefont
  {Xiong}}, \bibinfo {author} {\bibfnamefont {S.~K.}\ \bibnamefont {Kushwaha}},
  \bibinfo {author} {\bibfnamefont {T.}~\bibnamefont {Liang}}, \bibinfo
  {author} {\bibfnamefont {J.~W.}\ \bibnamefont {Krizan}}, \bibinfo {author}
  {\bibfnamefont {M.}~\bibnamefont {Hirschberger}}, \bibinfo {author}
  {\bibfnamefont {W.}~\bibnamefont {Wang}}, \bibinfo {author} {\bibfnamefont
  {R.~J.}\ \bibnamefont {Cava}}, \ and\ \bibinfo {author} {\bibfnamefont
  {N.~P.}\ \bibnamefont {Ong}},\ }\href {\doibase 10.1126/science.aac6089}
  {\bibfield  {journal} {\bibinfo  {journal} {Science}\ }\textbf {\bibinfo
  {volume} {350}},\ \bibinfo {pages} {413} (\bibinfo {year}
  {2015})}\BibitemShut {NoStop}%
\bibitem [{\citenamefont {Meng}\ and\ \citenamefont {Balents}(2012)}]{Meng12}%
  \BibitemOpen
  \bibfield  {author} {\bibinfo {author} {\bibfnamefont {T.}~\bibnamefont
  {Meng}}\ and\ \bibinfo {author} {\bibfnamefont {L.}~\bibnamefont {Balents}},\
  }\href {\doibase 10.1103/PhysRevB.86.054504} {\bibfield  {journal} {\bibinfo
  {journal} {Phys. Rev. B}\ }\textbf {\bibinfo {volume} {86}},\ \bibinfo
  {pages} {054504} (\bibinfo {year} {2012})}\BibitemShut {NoStop}%
\bibitem [{\citenamefont {Cho}\ \emph {et~al.}(2012)\citenamefont {Cho},
  \citenamefont {Bardarson}, \citenamefont {Lu},\ and\ \citenamefont
  {Moore}}]{Moore12}%
  \BibitemOpen
  \bibfield  {author} {\bibinfo {author} {\bibfnamefont {G.~Y.}\ \bibnamefont
  {Cho}}, \bibinfo {author} {\bibfnamefont {J.~H.}\ \bibnamefont {Bardarson}},
  \bibinfo {author} {\bibfnamefont {Y.-M.}\ \bibnamefont {Lu}}, \ and\ \bibinfo
  {author} {\bibfnamefont {J.~E.}\ \bibnamefont {Moore}},\ }\href {\doibase
  10.1103/PhysRevB.86.214514} {\bibfield  {journal} {\bibinfo  {journal} {Phys.
  Rev. B}\ }\textbf {\bibinfo {volume} {86}},\ \bibinfo {pages} {214514}
  (\bibinfo {year} {2012})}\BibitemShut {NoStop}%
\bibitem [{\citenamefont {Wei}\ \emph {et~al.}(2014)\citenamefont {Wei},
  \citenamefont {Chao},\ and\ \citenamefont {Aji}}]{Aji14}%
  \BibitemOpen
  \bibfield  {author} {\bibinfo {author} {\bibfnamefont {H.}~\bibnamefont
  {Wei}}, \bibinfo {author} {\bibfnamefont {S.-P.}\ \bibnamefont {Chao}}, \
  and\ \bibinfo {author} {\bibfnamefont {V.}~\bibnamefont {Aji}},\ }\href
  {\doibase 10.1103/PhysRevB.89.014506} {\bibfield  {journal} {\bibinfo
  {journal} {Phys. Rev. B}\ }\textbf {\bibinfo {volume} {89}},\ \bibinfo
  {pages} {014506} (\bibinfo {year} {2014})}\BibitemShut {NoStop}%
\bibitem [{\citenamefont {Lu}\ \emph {et~al.}(2015{\natexlab{b}})\citenamefont
  {Lu}, \citenamefont {Yada}, \citenamefont {Sato},\ and\ \citenamefont
  {Tanaka}}]{Tanaka15}%
  \BibitemOpen
  \bibfield  {author} {\bibinfo {author} {\bibfnamefont {B.}~\bibnamefont
  {Lu}}, \bibinfo {author} {\bibfnamefont {K.}~\bibnamefont {Yada}}, \bibinfo
  {author} {\bibfnamefont {M.}~\bibnamefont {Sato}}, \ and\ \bibinfo {author}
  {\bibfnamefont {Y.}~\bibnamefont {Tanaka}},\ }\href {\doibase
  10.1103/PhysRevLett.114.096804} {\bibfield  {journal} {\bibinfo  {journal}
  {Phys. Rev. Lett.}\ }\textbf {\bibinfo {volume} {114}},\ \bibinfo {pages}
  {096804} (\bibinfo {year} {2015}{\natexlab{b}})}\BibitemShut {NoStop}%
\bibitem [{\citenamefont {Bednik}\ \emph {et~al.}(2015)\citenamefont {Bednik},
  \citenamefont {Zyuzin},\ and\ \citenamefont {Burkov}}]{Bednik15}%
  \BibitemOpen
  \bibfield  {author} {\bibinfo {author} {\bibfnamefont {G.}~\bibnamefont
  {Bednik}}, \bibinfo {author} {\bibfnamefont {A.~A.}\ \bibnamefont {Zyuzin}},
  \ and\ \bibinfo {author} {\bibfnamefont {A.~A.}\ \bibnamefont {Burkov}},\
  }\href {\doibase 10.1103/PhysRevB.92.035153} {\bibfield  {journal} {\bibinfo
  {journal} {Phys. Rev. B}\ }\textbf {\bibinfo {volume} {92}},\ \bibinfo
  {pages} {035153} (\bibinfo {year} {2015})}\BibitemShut {NoStop}%
\bibitem [{\citenamefont {Yang}\ \emph {et~al.}(2014)\citenamefont {Yang},
  \citenamefont {Pan},\ and\ \citenamefont {Zhang}}]{FanZhang14}%
  \BibitemOpen
  \bibfield  {author} {\bibinfo {author} {\bibfnamefont {S.~A.}\ \bibnamefont
  {Yang}}, \bibinfo {author} {\bibfnamefont {H.}~\bibnamefont {Pan}}, \ and\
  \bibinfo {author} {\bibfnamefont {F.}~\bibnamefont {Zhang}},\ }\href
  {\doibase 10.1103/PhysRevLett.113.046401} {\bibfield  {journal} {\bibinfo
  {journal} {Phys. Rev. Lett.}\ }\textbf {\bibinfo {volume} {113}},\ \bibinfo
  {pages} {046401} (\bibinfo {year} {2014})}\BibitemShut {NoStop}%
\bibitem [{\citenamefont {Xu}\ \emph {et~al.}(2015{\natexlab{b}})\citenamefont
  {Xu}, \citenamefont {Zhang},\ and\ \citenamefont {Zhang}}]{FanZhang15}%
  \BibitemOpen
  \bibfield  {author} {\bibinfo {author} {\bibfnamefont {Y.}~\bibnamefont
  {Xu}}, \bibinfo {author} {\bibfnamefont {F.}~\bibnamefont {Zhang}}, \ and\
  \bibinfo {author} {\bibfnamefont {C.}~\bibnamefont {Zhang}},\ }\href
  {\doibase 10.1103/PhysRevLett.115.265304} {\bibfield  {journal} {\bibinfo
  {journal} {Phys. Rev. Lett.}\ }\textbf {\bibinfo {volume} {115}},\ \bibinfo
  {pages} {265304} (\bibinfo {year} {2015}{\natexlab{b}})}\BibitemShut
  {NoStop}%
\bibitem [{\citenamefont {{Li}}\ and\ \citenamefont
  {{Haldane}}(2015)}]{YiLi15}%
  \BibitemOpen
  \bibfield  {author} {\bibinfo {author} {\bibfnamefont {Y.}~\bibnamefont
  {{Li}}}\ and\ \bibinfo {author} {\bibfnamefont {F.~D.~M.}\ \bibnamefont
  {{Haldane}}},\ }\href@noop {} {\bibfield  {journal} {\bibinfo  {journal}
  {ArXiv e-prints}\ } (\bibinfo {year} {2015})},\ \Eprint
  {http://arxiv.org/abs/1510.01730} {arXiv:1510.01730 [cond-mat.str-el]}
  \BibitemShut {NoStop}%
\bibitem [{\citenamefont {Wang}\ \emph {et~al.}(2016)\citenamefont {Wang},
  \citenamefont {Hao}, \citenamefont {Wang},\ and\ \citenamefont
  {Ting}}]{Ting16}%
  \BibitemOpen
  \bibfield  {author} {\bibinfo {author} {\bibfnamefont {R.}~\bibnamefont
  {Wang}}, \bibinfo {author} {\bibfnamefont {L.}~\bibnamefont {Hao}}, \bibinfo
  {author} {\bibfnamefont {B.}~\bibnamefont {Wang}}, \ and\ \bibinfo {author}
  {\bibfnamefont {C.~S.}\ \bibnamefont {Ting}},\ }\href {\doibase
  10.1103/PhysRevB.93.184511} {\bibfield  {journal} {\bibinfo  {journal} {Phys.
  Rev. B}\ }\textbf {\bibinfo {volume} {93}},\ \bibinfo {pages} {184511}
  (\bibinfo {year} {2016})}\BibitemShut {NoStop}%
\bibitem [{\citenamefont {Murakami}\ and\ \citenamefont
  {Nagaosa}(2003)}]{Murakami03}%
  \BibitemOpen
  \bibfield  {author} {\bibinfo {author} {\bibfnamefont {S.}~\bibnamefont
  {Murakami}}\ and\ \bibinfo {author} {\bibfnamefont {N.}~\bibnamefont
  {Nagaosa}},\ }\href {\doibase 10.1103/PhysRevLett.90.057002} {\bibfield
  {journal} {\bibinfo  {journal} {Phys. Rev. Lett.}\ }\textbf {\bibinfo
  {volume} {90}},\ \bibinfo {pages} {057002} (\bibinfo {year}
  {2003})}\BibitemShut {NoStop}%
\bibitem [{\citenamefont {Yakovenko}(2007)}]{Yakovenko07}%
  \BibitemOpen
  \bibfield  {author} {\bibinfo {author} {\bibfnamefont {V.~M.}\ \bibnamefont
  {Yakovenko}},\ }\href {\doibase 10.1103/PhysRevLett.98.087003} {\bibfield
  {journal} {\bibinfo  {journal} {Phys. Rev. Lett.}\ }\textbf {\bibinfo
  {volume} {98}},\ \bibinfo {pages} {087003} (\bibinfo {year}
  {2007})}\BibitemShut {NoStop}%
\bibitem [{\citenamefont {Goryo}(2008)}]{Goryo08}%
  \BibitemOpen
  \bibfield  {author} {\bibinfo {author} {\bibfnamefont {J.}~\bibnamefont
  {Goryo}},\ }\href {\doibase 10.1103/PhysRevB.78.060501} {\bibfield  {journal}
  {\bibinfo  {journal} {Phys. Rev. B}\ }\textbf {\bibinfo {volume} {78}},\
  \bibinfo {pages} {060501} (\bibinfo {year} {2008})}\BibitemShut {NoStop}%
\bibitem [{\citenamefont {Lutchyn}\ \emph {et~al.}(2008)\citenamefont
  {Lutchyn}, \citenamefont {Nagornykh},\ and\ \citenamefont
  {Yakovenko}}]{Lutchyn08}%
  \BibitemOpen
  \bibfield  {author} {\bibinfo {author} {\bibfnamefont {R.~M.}\ \bibnamefont
  {Lutchyn}}, \bibinfo {author} {\bibfnamefont {P.}~\bibnamefont {Nagornykh}},
  \ and\ \bibinfo {author} {\bibfnamefont {V.~M.}\ \bibnamefont {Yakovenko}},\
  }\href {\doibase 10.1103/PhysRevB.77.144516} {\bibfield  {journal} {\bibinfo
  {journal} {Phys. Rev. B}\ }\textbf {\bibinfo {volume} {77}},\ \bibinfo
  {pages} {144516} (\bibinfo {year} {2008})}\BibitemShut {NoStop}%
\bibitem [{\citenamefont {Taylor}\ and\ \citenamefont
  {Kallin}(2012)}]{Kallin12}%
  \BibitemOpen
  \bibfield  {author} {\bibinfo {author} {\bibfnamefont {E.}~\bibnamefont
  {Taylor}}\ and\ \bibinfo {author} {\bibfnamefont {C.}~\bibnamefont
  {Kallin}},\ }\href {\doibase 10.1103/PhysRevLett.108.157001} {\bibfield
  {journal} {\bibinfo  {journal} {Phys. Rev. Lett.}\ }\textbf {\bibinfo
  {volume} {108}},\ \bibinfo {pages} {157001} (\bibinfo {year}
  {2012})}\BibitemShut {NoStop}%
\bibitem [{\citenamefont {Ojanen}\ and\ \citenamefont
  {Kitagawa}(2013)}]{Kitagawa13}%
  \BibitemOpen
  \bibfield  {author} {\bibinfo {author} {\bibfnamefont {T.}~\bibnamefont
  {Ojanen}}\ and\ \bibinfo {author} {\bibfnamefont {T.}~\bibnamefont
  {Kitagawa}},\ }\href {\doibase 10.1103/PhysRevB.87.014512} {\bibfield
  {journal} {\bibinfo  {journal} {Phys. Rev. B}\ }\textbf {\bibinfo {volume}
  {87}},\ \bibinfo {pages} {014512} (\bibinfo {year} {2013})}\BibitemShut
  {NoStop}%
\bibitem [{\citenamefont {Qi}\ \emph {et~al.}(2010)\citenamefont {Qi},
  \citenamefont {Hughes},\ and\ \citenamefont {Zhang}}]{Hughes10}%
  \BibitemOpen
  \bibfield  {author} {\bibinfo {author} {\bibfnamefont {X.-L.}\ \bibnamefont
  {Qi}}, \bibinfo {author} {\bibfnamefont {T.~L.}\ \bibnamefont {Hughes}}, \
  and\ \bibinfo {author} {\bibfnamefont {S.-C.}\ \bibnamefont {Zhang}},\ }\href
  {\doibase 10.1103/PhysRevB.82.184516} {\bibfield  {journal} {\bibinfo
  {journal} {Phys. Rev. B}\ }\textbf {\bibinfo {volume} {82}},\ \bibinfo
  {pages} {184516} (\bibinfo {year} {2010})}\BibitemShut {NoStop}%
\bibitem [{\citenamefont {Chung}\ and\ \citenamefont {Roy}(2014)}]{Roy14}%
  \BibitemOpen
  \bibfield  {author} {\bibinfo {author} {\bibfnamefont {S.~B.}\ \bibnamefont
  {Chung}}\ and\ \bibinfo {author} {\bibfnamefont {R.}~\bibnamefont {Roy}},\
  }\href {\doibase 10.1103/PhysRevB.90.224510} {\bibfield  {journal} {\bibinfo
  {journal} {Phys. Rev. B}\ }\textbf {\bibinfo {volume} {90}},\ \bibinfo
  {pages} {224510} (\bibinfo {year} {2014})}\BibitemShut {NoStop}%
\bibitem [{\citenamefont {Wang}\ \emph {et~al.}(2015)\citenamefont {Wang},
  \citenamefont {Zhou}, \citenamefont {Lian},\ and\ \citenamefont
  {Zhang}}]{Lian15}%
  \BibitemOpen
  \bibfield  {author} {\bibinfo {author} {\bibfnamefont {J.}~\bibnamefont
  {Wang}}, \bibinfo {author} {\bibfnamefont {Q.}~\bibnamefont {Zhou}}, \bibinfo
  {author} {\bibfnamefont {B.}~\bibnamefont {Lian}}, \ and\ \bibinfo {author}
  {\bibfnamefont {S.-C.}\ \bibnamefont {Zhang}},\ }\href {\doibase
  10.1103/PhysRevB.92.064520} {\bibfield  {journal} {\bibinfo  {journal} {Phys.
  Rev. B}\ }\textbf {\bibinfo {volume} {92}},\ \bibinfo {pages} {064520}
  (\bibinfo {year} {2015})}\BibitemShut {NoStop}%
\bibitem [{\citenamefont {Zyuzin}\ and\ \citenamefont
  {Burkov}(2012)}]{Zyuzin12-1}%
  \BibitemOpen
  \bibfield  {author} {\bibinfo {author} {\bibfnamefont {A.~A.}\ \bibnamefont
  {Zyuzin}}\ and\ \bibinfo {author} {\bibfnamefont {A.~A.}\ \bibnamefont
  {Burkov}},\ }\href {\doibase 10.1103/PhysRevB.86.115133} {\bibfield
  {journal} {\bibinfo  {journal} {Phys. Rev. B}\ }\textbf {\bibinfo {volume}
  {86}},\ \bibinfo {pages} {115133} (\bibinfo {year} {2012})}\BibitemShut
  {NoStop}%
\bibitem [{\citenamefont {Ramamurthy}\ and\ \citenamefont
  {Hughes}(2015)}]{Hughes15}%
  \BibitemOpen
  \bibfield  {author} {\bibinfo {author} {\bibfnamefont {S.~T.}\ \bibnamefont
  {Ramamurthy}}\ and\ \bibinfo {author} {\bibfnamefont {T.~L.}\ \bibnamefont
  {Hughes}},\ }\href {\doibase 10.1103/PhysRevB.92.085105} {\bibfield
  {journal} {\bibinfo  {journal} {Phys. Rev. B}\ }\textbf {\bibinfo {volume}
  {92}},\ \bibinfo {pages} {085105} (\bibinfo {year} {2015})}\BibitemShut
  {NoStop}%
\bibitem [{\citenamefont {Fu}\ and\ \citenamefont {Kane}(2007)}]{Fu-Kane}%
  \BibitemOpen
  \bibfield  {author} {\bibinfo {author} {\bibfnamefont {L.}~\bibnamefont
  {Fu}}\ and\ \bibinfo {author} {\bibfnamefont {C.~L.}\ \bibnamefont {Kane}},\
  }\href {\doibase 10.1103/PhysRevB.76.045302} {\bibfield  {journal} {\bibinfo
  {journal} {Phys. Rev. B}\ }\textbf {\bibinfo {volume} {76}},\ \bibinfo
  {pages} {045302} (\bibinfo {year} {2007})}\BibitemShut {NoStop}%
\end{thebibliography}%
\end{document}